\documentclass[11pt,preprint]{aastex}

\slugcomment{to appear with shortened appendicies in The Astrophysical Journal}

\shortauthors{Hardee}
\shorttitle{Strongly Magnetized Jet Stability}

\begin{document}
 
\baselineskip 12pt
\parskip 2pt

\author{Philip E. Hardee}
\affil{Department of Physics \& Astronomy, The University of Alabama,
Tuscaloosa, AL 35487, USA}
\email{phardee@bama.ua.edu}

\title{Stability Properties of Strongly Magnetized Spine Sheath
Relativistic Jets}

\begin{abstract}

\baselineskip 12pt
\parskip 2pt

The linearized relativistic magnetohydrodynamic (RMHD) equations
describing a uniform axially magnetized cylindrical relativistic jet spine
embedded in a uniform axially magnetized relativistically moving
sheath are derived. The displacement current is retained in the
equations so that effects associated with Alfv\'{e}n wave propagation
near light speed can be studied. A dispersion relation for the normal
modes is obtained. Analytical solutions for the normal modes in the
low and high frequency limits are found and a general stability
condition is determined. A trans-Alfv\'enic and even a super-Alf\'enic
relativistic jet spine can be stable to velocity shear driven
Kelvin-Helmholtz modes. The resonance condition for maximum growth
of the normal modes is obtained in the kinetically and magnetically
dominated regimes. Numerical solution of the dispersion relation
verifies the analytical solutions and is used to study the regime of high
sound and Alfv\'{e}n speeds.
  
\end{abstract}

\keywords{galaxies: jets --- gamma rays: bursts --- ISM: jets and
  outflows --- methods: analytical ---  MHD --- relativity --- instabilities
 \vspace{-0.5cm}}

\section{Introduction}

Relativistic jets are associated with active galactic nuclei and
quasars ({\bf AGN}), with black hole binary star systems ({\bf
microquasars}), and are thought responsible for the gamma-ray bursts
({\bf GRBs}).  In microquasar and AGN jets proper motions of intensity
enhancements show mildly superluminal for the microquasar jets $\sim
1.2~c$ (Mirabel \& Rodriquez 1999), range from subluminal ($\ll c$) to
superluminal ($\lesssim 6~c$) along the M\,87 jet (Biretta et al.
1995, 1999), are up to $\sim 25~c$ along the 3C\,345 jet (Zensus et
al.\ 1995; Steffen et al.\ 1995), and have inferred Lorentz factors
$\gamma > 100$ in the GRBs (e.g., Piran 2005).  The observed proper
motions along microquasar and AGN jets imply speeds from $\sim 0.9~c$
up to $\sim 0.999~c$, and the speeds inferred for the GRBs are $\sim
0.99999~c$.

Jets at the larger scales may be kinetically dominated and contain
relatively weak magnetic fields, e.g., equipartition between magnetic
and gas pressure or less, but the possibility of much stronger
magnetic fields exists close to the acceleration and collimation
region. Here general relativistic magnetohydrodynamic ({\bf GRMHD})
simulations of jet formation (e.g., Koide et al.\ 2000; Nishikawa et
al.\ 2005; De Villiers, Hawley \& Krolik 2003; De Villiers et al.\
2005; Hawley \& Krolik 2006; McKinney 2006; Mizuno et al.\ 2006) and
earlier theoretical work (e.g., Lovelace 1976; Blandford 1976;
Blandford \& Znajek 1977; Blandford \& Payne 1982) invoke strong
magnetic fields. In addition to strong magnetic fields, GRMHD
simulation studies of jet formation indicate that highly collimated
high speed jets driven by the magnetic fields threading the ergosphere
may themselves reside within a broader wind or sheath outflow driven
by the magnetic fields anchored in the accretion disk (e.g., McKinney
2006; Hawley \& Krolik 2006; Mizuno et al.\ 2006). This configuration
might additionally be surrounded by a less collimated accretion disk
wind from the hot corona (e.g., Nishikawa et al.\ 2005).

That relativistic jets may have jet-wind structure is indicated by
recent observations of high speed winds in several QSO's with speeds,
$\sim 0.1 - 0.4c$, (Chartas, Brandt \& Gallagher 2002, Chartas et al.\
2003; Pounds et al.\ 2003a; Pounds et al.\ 2003b; Reeves, O'Brien \&
Ward 2003).  Other observational evidence such as {\it limb
brightening} has been interpreted as evidence for a slower external
sheath flow surrounding a faster jet spine, e.g., Mkn\,501 (Giroletti
et al.\ 2004), M\,87 (Perlman et al.\ 2001), and a few other radio
galaxy jets (e.g., Swain, Bridle \& Baum 1998; Giovannini et al.\
2001).  Additional circumstantial evidence such as the requirement for
large Lorentz factors suggested by the TeV BL Lacs when contrasted
with much slower observed motions suggests the presence of a
spine-sheath morphology (Ghisellini, Tavecchio \& Chiaberge 2005).  At
hundreds of kiloparsec scales Siemignowska et al.\ (2007) have
proposed a two component (spine-sheath) model to explain the
broad-band emission from the PKS 1127-145 jet.  A spine-sheath jet
structure has been proposed based on theoretical arguments (e.g., Sol
et al.\ 1989; Henri \& Pelletier 1991; Laing 1996; Meier
2003). Similar type structure has been investigated in the context of
GRB jets (e.g., Rossi, Lazzati \& Rees 2002; Lazzatti \& Begelman
2005; Zhang, Wooseley \& MacFadyen 2003; Zhang, Woosley \& Heger 2004;
Morsony, Lazzati \& Begelman 2006).

In order to study the effect of strong magnetic fields and the effect
of a moving wind or sheath around a jet or jet spine, I begin by
adopting a simple system with no radial dependence of quantities
inside the jet spine and no radial dependence of quantities outside
the jet in the sheath.  This ``top hat'' configuration with magnetic
fields parallel to the flow can be described exactly by the linearized
relativistic magnetohydrodynamic ({\bf RMHD}) equations.  This system
with no magnetic and flow helicity is stable to current driven ({\bf
CD}) modes of instability (Istomin \& Pariev 1994, 1996; Lyubarskii
1999).  However, this system can be unstable to Kelvin-Helmholtz ({\bf
KH}) modes of instability (Hardee 2004). This approach allows us to
look at the potential KH modes without complications arising from
coexisting CD modes (see Baty, Keppens \& Compte 2004) and predictions
can be verified by numerical simulations (Mizuno, Hardee \& Nishikawa 2006).

This paper is organized as follows. In \S 2, I present the dispersion
relation arising from a normal mode analysis of the linearized RMHD
equations. Analytical approximate solutions to the dispersion relation
for various limiting cases are given in \S 3. I verify the analytical
solution through numerical solution of the dispersion relation in \S
4. I summarize the stability results in \S 5 and discuss the
applicability of the present results in \S 6.  Derivation of the
linearized RMHD equations is shown in Appendix A, derivation of the
normal mode dispersion relation is presented in Appendix B, and
derivation of the analytical solutions is shown in Appendix C.

\vspace{-0.5cm}
\section{The RMHD Normal Mode Dispersion Relation}

Let us analyze the stability of a spine-sheath system by modeling the
jet spine as a cylinder of radius R, having a uniform proper density,
$\rho _{j}$, a uniform axial magnetic field, $B_{j}=B_{j,z}$, and a
uniform velocity, $\mathbf{u}_{j}=u_{j,z}$.  The external sheath is
assumed to have a uniform proper density, $ \rho _{e}$, a uniform
axial magnetic field, $B_{e}=B_{e,z}$, a uniform velocity
$\mathbf{u}_{e}=u_{e,z}$, and extends to infinity. The sheath velocity
corresponds to an outflow around the central spine if $u_{e,z}>0$ or
represents backflow when $u_{e,z}<0$. The jet spine is established in
static total pressure balance with the external sheath where the total
static uniform pressure is $P_{e}^{\ast }\equiv P_{e}+B_{e}^{2}/8\pi
=P_{j}^{\ast }\equiv P_{j}+B_{j}^{2}/8\pi $, and the initial
equilibrium satisfies the zeroth order equations. Formally, the
assumption of an infinite sheath means that a dispersion relation
could be derived in the reference frame of the sheath with results
transformed to the source/observer reference frame. However, it is not
much more difficult to derive a dispersion relation in the
source/observer frame in which analytical solutions to the dispersion
relation take on simple revealing forms.  Additionally, this approach
lends itself to modeling the propagation and appearance of jet
structures viewed in the source/observer frame, e.g., helical
structures in the 3C\,120 jet (Hardee, Walker \& G\'omez 2005).

The general approach to analyzing the time dependent properties of
this system is to linearize the ideal RMHD and Maxwell equations,
where the density, velocity, pressure and magnetic field are written
as $\rho =\rho _{0}+\rho _{1}$, $\mathbf{v}=\mathbf{u}+\mathbf{v}_{1}$
(we use $\mathbf{v}_{0}\equiv \mathbf{u}$ for notational reasons),
$P=P_{0}+P_{1}$, and $\mathbf{B}= \mathbf{B}_{0}+\mathbf{B}_{1}$,
where subscript 1 refers to a perturbation to the equilibrium quantity
with subscript 0. Additionally, the Lorentz factor $\gamma
^{2}=(\gamma _{0}+\gamma _{1})^{2}\simeq \gamma _{0}^{2}+2\gamma
_{0}^{4}\mathbf{u\cdot v}_{1}/c^{2}$ where $\gamma _{1}=\gamma
_{0}^{3}\mathbf{u\cdot v}_{1}/c^{2}$. The linearization is shown in
Appendix A.  In cylindrical geometry a random perturbation $\rho
_{1}$, $\mathbf{v}_{1}$ $ \mathbf{B}_{1}$ and $P_{1}$ can be
considered to consist of Fourier components of the form
\begin{equation}
f_{1}(r,\phi ,z,t)=f_{1}(r)\exp [i(kz\pm n\phi -\omega t)]  
\label{1}
\end{equation}
where flow is along the z-axis, and r is in the radial direction with
the flow bounded by $r=R$. In cylindrical geometry $n$ is an integer
azimuthal wavenumber, for $n>0$ waves propagate at an angle to the flow
direction, and $+n$ and $-n$ give wave propagation in the clockwise and
counter-clockwise sense, respectively, when viewed in the flow direction. In
equation (1) $n=0$, $1$, $2$, $3$, $4$, etc. correspond to pinching,
helical, elliptical, triangular, rectangular, etc. normal mode distortions
of the jet, respectively. Propagation and growth or damping of the Fourier
components can be described by a dispersion relation of the form 
\begin{equation}
\frac{\beta _{j}}{\chi _{j}}\frac{J_{n}^{^{\prime }}(\beta _{j}R)}{
J_{n}(\beta _{j}R)}=\frac{\beta _{e}}{\chi _{e}}\frac{H_{n}^{(1)^{\prime
}}(\beta _{e}R)}{H_{n}^{(1)}(\beta _{e}R)}~.  
\label{2}
\end{equation}
Derivation of this dispersion relation is given in Appendix B.
In the dispersion relation $J_{n}$ and $H_{n}^{(1)}$ are Bessel and Hankel
functions, the primes denote derivatives of the Bessel and Hankel functions
with respect to their arguments. In equation (2)
\begin{equation}
\eqnum{3a}
\chi _{j}\equiv \gamma _{j}^{2}\gamma _{Aj}^{2}W_{j}\left( \varpi
_{j}^{2}-\kappa _{j}^{2}v_{Aj}^{2}\right)~,  
\label{3a}
\end{equation}
\begin{equation}
\eqnum{3b}
\chi _{e}\equiv \gamma _{e}^{2}\gamma _{Ae}^{2}W_{e}\left( \varpi
_{e}^{2}-\kappa _{e}^{2}v_{Ae}^{2}\right)~,  
\label{3b}
\end{equation}
\setcounter{equation}{3}
and
\begin{equation}
\eqnum{4a}
\beta _{j}^{2}\equiv \left[ \frac{\gamma _{j}^{2}\left( \varpi
_{j}^{2}-\kappa _{j}^{2}a_{j}^{2}\right) \left( \varpi _{j}^{2}-\kappa
_{j}^{2}v_{Aj}^{2}\right) }{v_{msj}^{2} \varpi _{j}^{2}-\kappa
_{j}^{2}v_{Aj}^{2}a_{j}^{2}}\right]~,
\label{4a}
\end{equation}
\begin{equation}
\eqnum{4b}
\beta _{e}^{2}\equiv \left[ \frac{\gamma _{e}^{2}\left( \varpi
_{ex}^{2}-\kappa _{e}^{2}a_{e}^{2}\right) \left( \varpi
_{e}^{2}-\kappa _{e}^{2}v_{Ae}^{2}\right) }{v_{mse}^{2} \varpi
_{e}^{2}-\kappa _{e}^{2}v_{Ae}^{2}a_{e}^{2}}\right]~.
\label{4b}
\end{equation}
\setcounter{equation}{4}
In equations (3a \& 3b) and equations (4a \& 4b) $\varpi
_{j,e}^{2}\equiv \left( \omega -ku_{j,e}\right) ^{2}$ and $\kappa
_{j,e}^{2}\equiv \left( k-\omega u_{j,e}/c^{2}\right) ^{2}$, $\gamma
_{j,e}\equiv (1-u_{j,e}^{2}/c^{2})^{-1/2} $ is the flow Lorentz
factor, $\gamma _{Aj,e}\equiv (1-v_{Aj,e}^{2}/c^{2})^{-1/2}$ is the
Alfv\'{e}n Lorentz factor, $W\equiv \rho +\left[ \Gamma /\left( \Gamma
-1\right) \right] P/c^{2} $ is the enthalpy, $a$ is the sound speed,
$v_{A}$ is the Alfv\'{e}n wave speed, and $v_{ms}$ is a magnetosonic
speed. The sound speed is defined by
$$
a\equiv \left[ \frac{\Gamma P}{\rho +\left( \frac{\Gamma }{\Gamma -1}\right)
P/c^{2}}\right] ^{1/2}~,
$$
where $4/3\leq \Gamma \leq 5/3$ is the adiabatic index. The
Alfv\'{e}n wave speed is defined by
$$
v_{A}\equiv \left[ \frac{V_{A}^{2}}{1+V_{A}^{2}/c^{2}}\right] ^{1/2}
$$
where $V_{A}^{2}\equiv B_{0}^{2}/(4\pi W_{0})$. A magnetosonic speed
corresponding to the fast magnetosonic speed for propagation
perpendicular to the magnetic field (e.g., Vlahakis \& K\"onigl 2003)
is defined by
$$
v_{ms} \equiv \left[a^2 + v_A^2 - a^2v_A^2/c^2\right]^{1/2} =
\left[a^2/\gamma_A^2 + v_A^2\right]^{1/2}~.
$$

\vspace{-0.5cm}
\section{Analytical Solutions to the Dispersion Relation}

In this section analytical solutions to the dispersion relation in the
low frequency limit, in the fluid and magnetic limits at resonance
(maximum growth), and in the high frequency limit are summarized. The
analytical solutions are derived in Appendix C.

\vspace{-0.5cm}
\subsection{Low Frequency Limit}

Analytically each normal mode $n$ contains a single
\textit{fundamental/surface} wave ($\omega \longrightarrow 0$,
$k\longrightarrow 0$, $\omega /k>0$) solution and multiple
\textit{body wave }($\omega \longrightarrow 0$, $k>0$, $\omega
/k\longrightarrow 0$) solutions that satisfy the dispersion
relation. In the low frequency limit the \textit{fundamental } pinch
mode ($n=0$) solution is given by
\begin{equation}
\frac{\omega }{k}=\frac{u_{j}\pm v_{w}}{1\pm v_{w}u_{j}/c^{2}}  
\label{5}
\end{equation}
where the  pinch fundamental mode wave speed
\begin{equation}
v_{w}^{2}\approx a_{j}^{2}\left\{ \frac{v_{Aj}^{2}}{
v_{msj}^{2} }+\delta \left[ \frac{v_{Aj}^{2}}{v_{msj}^{2} }-1
\right] \right\}~,  
\label{6}
\end{equation}
and
\begin{equation}
\delta \equiv -\frac{1}{2}\gamma _{e}^{2}\frac{\gamma
_{Ae}^{2}}{\gamma_{Aj}^2}\frac{W_{e}}{W_{j}} \frac{\left( \varpi
_{e}^{2}-\kappa _{e}^{2}v_{Ae}^{2}\right) }{ v_{msj}^{2} }R^{2}\left[
\ln (\frac{\beta _{e}R}{2})+\frac{\pi }{2}\epsilon -i\frac{\pi
}{2}\right]
\label{7}
\end{equation}
with $\left| \delta \right| \propto \left| k^{2}R^{2}\right| <<1$. \
In this limit $\delta$ is complex and this mode consists of a growing
and damped wave pair. The imaginary part of the solution is
vanishingly small in the low frequency limit. The above form indicates
that growth, which arises from the complex value of $\delta $, will be
reduced as $ ( v_{Aj}^{2}/v_{msj}^{2}) \longrightarrow 1$. The
unstable growing solution is associated with the backwards moving (in
the jet fluid reference frame) wave.

In the low frequency limit the \textit{surface} helical, elliptical,
and higher order normal modes ($n>0$) have a solution given by
\begin{equation}
\frac{\omega }{k}=\frac{\left[ \eta u_{j}+u_{e}\right] \pm i\eta
^{1/2}\left[ \left( u_{j}-u_{e}\right) ^{2}-V_{As}^{2}/\gamma
_{j}^{2}\gamma _{e}^{2} \right] ^{1/2}}{(1+V_{Ae}^{2}/\gamma
_{e}^{2}c^{2})+\eta (1+V_{Aj}^{2}/\gamma _{j}^{2}c^{2})}
\label{8}
\end{equation}
where $\eta \equiv \gamma _{j}^{2}W_{j}\left/ \gamma _{e}^{2}W_{e}\right. $
and a ``surface'' Alfv\'{e}n speed is defined by
\begin{equation}
V_{As}^{2}\equiv \left( \gamma _{Aj}^{2}W_{j}+\gamma _{Ae}^{2}W_{e}\right) 
\frac{B_{j}^{2}+B_{e}^{2}}{4\pi W_{j}W_{e}}~.  
\label{9}
\end{equation}
In equation (9) note that the Alfv\'{e}n Lorentz factor $\gamma
_{Aj,e}^{2}=1+V_{Aj,e}^{2}/c^{2}$. Thus, the jet is stable to $n>0$
surface wave mode perturbations when
\begin{equation}
\gamma _{j}^{2}\gamma _{e}^{2}\left( u_{j}-u_{e}\right)^{2} < \gamma
_{Aj}^{2}\gamma _{Ae}^{2}\left( W_{j}/\gamma _{Ae}^{2}+W_{e}/\gamma
_{Aj}^{2}\right) \frac{B_{j}^{2}+B_{e}^{2}}{4\pi W_{j}W_{e}}~.
\label{10}
\end{equation}
For example, with $u_{j}\approx c>>u_{e}$, $\gamma_{e}^{2} \approx 1$,
$ \gamma _{Aj}^{2} >> \gamma_{Ae}^{2}\approx 1$,
$B_{j}^{2}>>B_{e}^{2}$, and using $\gamma _{Aj}^{2}=1+B_{j}^{2}/4\pi
W_{j}c^{2}$ the jet is stable when
\begin{equation} \eqnum{11a}
\gamma _{j}^{2} < \left[ 1+\frac{B_{j}^{2}}{4\pi
    W_{e}c^{2}}\right]\gamma _{Aj}^{2}
\label{11a}
\end{equation}
or with $B_e = B_j$, $W_e = W_j$, so that $v_{A,j} = v_{A,e}$, and
with $\gamma_A \equiv \gamma_{A,e} = \gamma_{A,j}$ the jet is stable when
\begin{equation} \eqnum{11b}
\gamma _{j}^{2} \gamma _{e}^{2} (u_j - u_e)^2 < 4 \gamma _{A}^{2}
(\gamma _{A}^{2} - 1) c^2~.
\label{11b}
\end{equation}
\setcounter{equation}{11}
Thus, the jet can remain stable to the surface wave modes even when the jet
Lorentz factor exceeds the Alfv\'{e}n Lorentz factor.

In the low frequency limit the real part of the \textit{body wave} solutions
is given by
\begin{equation}
kR\approx k_{nm}^{\min }R\equiv \left[ \frac{
v_{msj}^2u_{j}^{2}-v_{Aj}^2a_{j}^{2}}{\gamma
_{j}^{2}(u_{j}^{2}-a_{j}^{2})(u_{j}^{2}-v_{Aj}^{2})}\right]
^{1/2}\times \left[ (n+2m-1/2)\pi /2+(-1)^{m}\epsilon _{n}\right]
\label{12}
\end{equation}
where $n$ specifies the normal mode, $m=1,2,3,...$ specify the first,
second, third, etc. body wave solutions, and
$$
\epsilon _{n}\equiv \frac{\chi _{e}}{\chi _{j}}\frac{\beta _{e}}{\beta_{j}}
\left( \frac{\pi \beta _{j}R}{2}\right) ^{1/2}J_{n}^{^{\prime }}(\beta _{j}R)
\frac{H_{n}^{(1)}(\beta _{e}R)}{H_{n}^{(1)^{\prime }}(\beta _{e}R)}~.
$$
In the absence of a significant external magnetic field and a
significant external flow $\epsilon _{n}=0$ as $\chi
_{e}=\gamma _{e}^{2}\gamma _{Ae}^{2}W_{e}\left[ u_{e}^{2}-v_{Ae}^{2}\right]
k^{2}=0$. In this low frequency limit the body wave solutions are
either purely real or damped, exist only when $k_{nm}^{\min }R$ has a
positive real part, and with $\left| \epsilon _{n}\right| <<1$ require
that
\begin{equation}
 \left[ \frac{ v_{msj}^2u_{j}^{2}-v_{Aj}^2a_{j}^{2}}{\gamma
 _{j}^{2}(u_{j}^{2}-a_{j}^{2})(u_{j}^{2}-v_{Aj}^{2})}\right] >0~.
\label{13}
\end{equation}
Thus, the body modes can exist when the jet is supersonic and
super-Alfv\'{e}nic, i.e., $u_{j}^{2}-a_{j}^{2}>0$ and
$u_{j}^{2}-v_{Aj}^{2}>0$, or in a limited velocity range given
approximately by $a_{j}^{2}>u_{j}^{2}>[\gamma _{sj}^{2}/(1+\gamma
_{sj}^{2})]a_{j}^{2}$ when $v_{Aj}^{2}\approx a_{j}^{2}$, where
$\gamma _{sj} \equiv (1-a_{j}^2/c^2)^{-1/2}$ is a sonic Lorentz
factor.

\vspace{-0.5cm}
\subsection{Resonance}

With the exception of the pinch fundamental mode which can have a
relatively broad plateau in the growth rate, all \textit{body} modes,
and all \textit{surface} modes can have a distinct maximum in the
growth rate at some resonant frequency.

The resonance condition can be evaluated analytically in either the
fluid limit where $a>>V_{A}$ or in the magnetic limit where
$V_{A}>>a$. Note that in the magnetic limit, magnetic pressure balance
implies that $B_j = B_e$. In these cases a necesary condition for
resonance is that
\begin{equation}
\frac{u_{j}-u_{e}}{1-u_{j}u_{e}/c^{2}}>\frac{v_{wj}+v_{we}}{
1+v_{wj}v_{we}/c^{2}}~,  
\label{14}
\end{equation}
where $v_{wj}\equiv \left( a_{j},v_{Aj}\right) $ and $v_{we}\equiv \left(
a_{e},v_{Ae}\right) $ in the fluid or magnetic limits, respectively. When
this condition is satisfied it can be shown that the wave speed at
resonance is
\begin{equation}
v_{w} \approx v_{w}^{\ast }\equiv \frac{\gamma _{j}(\gamma
_{we}v_{we})u_{j}+\gamma _{e}(\gamma _{wj}v_{wj})u_{e}}{\gamma
_{j}(\gamma _{we}v_{we})+\gamma _{e}(\gamma _{wj}v_{wj})}
\label{15}
\end{equation}
where $\gamma _{w}\equiv (1-v_{w}^{2}/c^{2})^{-1/2}$ is the sonic or
Alfv\'{e}nic Lorentz factor accompanying $v_{wj}\equiv \left(
a_{j},v_{Aj}\right) $ and $v_{we}\equiv \left( a_{e},v_{Ae}\right) $ in the
fluid or magnetic limits, respectively.  

The resonant wave speed and maximum growth rate occur at a frequency
given by
\begin{equation}
\omega R/v_{we} \approx \omega_{nm}^{\ast}R/v_{we} \equiv
\frac{(2n+1)\pi /4+m\pi }{ \left[\left(1 - u_{e}/v^{\ast}_w\right)^2 -
\left(v_{we}/v^{\ast}_w - u_e v_{we}/c^2\right)^2\right]^{1/2}}~.
\label{16}
\end{equation}
In equation (16) $n$ specifies the normal mode, $m=0$ specifies the
surface wave, and $m\geq 1$ specifies the body waves.  In the limit of
insignificant sheath flow, $u_e = 0$, and using eq.\ (15) for
$v_w^{\ast}$ in eq.\ (16) allows the resonant frequency to be written as
$$
\omega_{nm}^{\ast} R_j/v_{we} = \frac{(2n+1)\pi /4+m\pi }{ \left[1-
\left(v_{we}^{2}/u_{j}^{2}+2 \frac{\gamma_{wj}}{\gamma_{we}}v_{we}
v_{wj}/\gamma_j u_j^2+\frac{\gamma
_{wj}^2}{\gamma_{we}^2}v_{wj}^2/\gamma_j^2u_j^2\right)\right]^{1/2}}~,
$$
and this predicts a resonant frequency that is primarily a function of
the sound and Alfv\'en wave speeds in the sheath. The effect of sheath
flow is best illustrated by assuming comparable conditions in the
spine and sheath, $\gamma_{wj}v_{wj} \sim \gamma_{we}v_{we}$, and
assuming that $\gamma_ju_j >> \gamma_eu_e$ in which case
$$
\omega_{nm}^{\ast} R_j/v_{we} \sim \frac{(2n+1)\pi /4+m\pi }{\gamma_e
\left[1- 2\left(u_e/u_j\right)(1-v_{we}^{2}/c^{2}) -
\left(v_{we}^2-u_e^2\right)/u_j^2\right]^{1/2}} ~.
$$
The term $u_e/u_j$ in the denominator indicates that the resonant
frequency increases as the shear speed, $u_j - u_e$, declines. In the
limit
$$
\frac{u_{j}-u_{e}}{1-u_{j}u_{e}/c^{2}}\longrightarrow \frac{v_{wj}+v_{we}}{
1+v_{wj}v_{we}/c^{2}}~,
$$
the resonant frequency $\omega _{nm}^{\ast }R/v_{we}\longrightarrow
\infty $.

The resonant wavelength is given by $\lambda \approx \lambda
_{nm}^{\ast }\equiv 2\pi v_{w}^{\ast }/\omega_{nm}^{\ast }$ and can be
calculated from
\begin{equation}
\lambda _{nm}^{\ast }\equiv\frac{2\pi}{(2n+1)\pi
/4+m\pi}\left(\frac{\gamma_e}{v_{we}}\right)\left\{\left(v_{w}^{\ast}
- u_e\right)^2 - \left[v_{we} -
(v_{we}u_e/c^2)v_{w}^{\ast}\right]^2\right\}^{1/2} R ~.
\label{17}
\end{equation} 
Equations (15 - 17) provide the proper functional dependence of the resonant
wave speed, frequency and wavelength provided $(u_e/u_j)^2 << 1$ and
$(v_{we}/u_j)^2 << 1$.

With the exception of the $n=0$, $m=0$, fundamental pinch mode, a
maximum spatial growth rate, $k_{I}^{\rm max}$, is approximated by
\begin{equation}
k_{I}^{\rm max}R \approx k_{I}^{\ast}R \equiv
-\frac{1}{2}\frac{v_{wj}}{\gamma _{j}u_{j}}\ln \left| 
\mathcal{R}\right| ~,  
\label{18}
\end{equation}
where
\begin{equation}
\left| \mathcal{R}\right| \approx \left[ \frac{4\left(\omega
_{nm}^{\ast }R/v_{we}\right) ^{2}\left( 1-2u_{e}/u_{j}\right) +\left( \ln
\left| \mathcal{R}\right| /2\right) ^{2}}{\left( \ln \left| \mathcal{R}
\right| /2\right) ^{2}}\right] ^{1/2} ~.  
\label{19}
\end{equation}
Equations (18) and (19) show that the maximum growth rate is primarily
a function of the jet sound, Alfv\'en and flow speed through
$v_{wj}/\gamma _{j}u_{j}$, and secondarily a function of the sheath
sound, Alfv\'en and flow speed through $\left(\omega
_{nm}^{\ast }R/v_{we}\right)^{2}\left( 1-2u_{e}/u_{j}\right)$.

I can illustrate the dependencies of the maximum growth rate on sound,
Alfv\'en and flow speeds by using
$$
\left( \frac{\omega _{nm}^{\ast }R}{v_{we}}\right) ^{2}\left(
1-2u_{e}/u_{j}\right) \approx \frac{\left( 1-2u_{e}/u_{j}\right)
}{\left[ 1-2\left( u_{e}/u_{j}\right)
(1-v_{we}^{2}/c^{2})-(v_{we}^{2}-u_e^2)/u_{j}^{2}\right] } \times
\left[ (2n+1)\pi /4+m\pi \right] ^{2}
$$
and if say $u_{e}=0$, then
\begin{equation}
\left( \left| \mathcal{R}\right| ^{2}-1\right) ^{1/2}\ln \left| \mathcal{R}
\right| \approx 4\left[ \frac{1}{1-v_{we}^{2}/u_{j}^{2}}\right] ^{1/2}
\times \left[
(2n+1)\pi /4+m\pi \right]~.  
\label{20}
\end{equation}
Thus, $\left| \mathcal{R}\right| $ increases as $\omega _{nm}^{\ast }$
increases for higher order modes with larger $n$ and larger $m$ and
this result indicates an increase in the growth rate for larger $n$
and larger $m$.  When the sound or Alfv\'{e}n wave speed, $v_{we}$,
increases $\left| \mathcal{R}\right| $
increases. This result indicates an increase in the growth rate at the
higher resonant frequency accompanying an increase in the sound or
Alfv\'{e}n wave speed in the sheath.

The behavior of the maximum growth rate as the shear speed, $u_j -
u_e$, declines is best illustrated by considering the effect of an
increasing wind speed where $(v_{we}^{2}-u_e^2)/u_{j}^{2}<<1$ is
ignored. In this case
\begin{equation}
\left( \left| \mathcal{R}\right| ^{2}-1\right) ^{1/2}\ln \left| \mathcal{R}
\right| \approx 4\left[ (2n+1)\pi /4+m\pi \right]  
\label{21}
\end{equation}
and $\left| \mathcal{R}\right| $ will remain relatively independent of
$\omega_{nm}^{\ast}$ even as $\omega _{nm}^{\ast }\longrightarrow
\infty $ as the shear speed decreases. This result indicates a
relatively constant resonant growth rate as the shear speed
decreases. 

In the fluid limit decline in the shear speed ultimately results in a
decrease in the growth rate and increase in the spatial growth length.
This decline in the growth rate is also indicated by equation (8)
which, in the fluid limit, becomes
\begin{equation}
\frac{\omega }{k}=\frac{\eta u_{j}+u_{e}}{1+\eta }\pm i\frac{\eta ^{1/2}}{
1+\eta }\left( u_{j}-u_{e}\right)~.  
\label{22}
\end{equation}
Equation (22) applies to frequencies below the resonant frequency
$\omega _{nm}^{\ast }$ and directly reveals the decline in growth
rates as $ u_{j}-u_{e}\longrightarrow 0$.

In the magnetic limit the resonant frequency $\omega _{nm}^{\ast
}R/v_{Ae} \longrightarrow \infty $ as
\begin{equation}
\frac{u_{j}-u_{e}}{1-u_{j}u_{e}/c^{2}} \longrightarrow \frac{v_{Aj}+v_{Ae}}{
1+v_{Aj}v_{Ae}/c^{2}}~.  \label{23}
\end{equation}
Here equation (8) indicates that the jet is stable when
$$
\gamma_{j}^{2}\gamma_{e}^{2}\left(u_{j} - u_{e}\right)^{2} < V_{As}^{2}~,
$$
and the jet will be stable as $\omega_{nm}^{\ast} \longrightarrow \infty$ when
\begin{equation}
\gamma _{j}^{2}\gamma _{e}^{2}\left( 1-u_{j}u_{e}/c^{2}\right) ^{2} < 2\gamma
_{Aj}^{2}\gamma _{Ae}^{2}\frac{v_{Ae}^{2}+v_{Aj}^{2}}{\left(
v_{Aj}+v_{Ae}\right) ^{2}}\left( 1+v_{Aj}v_{Ae}/c^{2}\right) ^{2}~,
\label{24}
\end{equation}
where I have used an equality in equation (23) in equation (8) to
obtain equation (24).  Equation (24) indicates that a high jet speed
relative to the Alfv\'{e}n wave speed is necessary for instability.
For example, if $v_{A} \equiv v_{Aj}= v_{Ae}$ and $u_{e} = 0$, the jet
is stable at high frequencies provided
\begin{equation} 
\eqnum{25a}
\gamma_{j}^2 < \left( 1 + v_{A}^{2}/c^{2} \right)^2 \gamma_{A}^{4}~.
\label{25a}
\end{equation}
This high frequency condition is slightly different from the low frequency
stabilization condition found when $v_{A} \equiv v_{Aj}= v_{Ae}$ and
$u_{e} = 0$ from equation (11b)
\begin{equation} \eqnum{25b}
\gamma_{j}^2 (u_j/c)^2 < 4 \gamma_{A}^{2}(\gamma_{A}^{2} - 1)~.
\label{25b}
\end{equation}
\setcounter{equation}{25}
Note that eqs.\ (25a \& 25b) are identical in the large Lorentz factor
limit.  Equations (25) predict that stabilization at high frequencies
occurs at somewhat higher jet speeds than stabilization at lower
frequencies. Determination of stabilization at intermediate
frequencies requires numerical solution of the dispersion relation. A
non-negligable postive external flow requires even higher jet speeds
for the jet to be unstable. Thus, a strongly magnetized relativistic
trans-Alfv\'{e}nic jet is predicted to be KH stable and a
super-Alfv\'enic jet can be KH stable.

\vspace{-0.5cm}
\subsection{High Frequency Limit}

Provided the condition, eq.\ (14), for resonance is met, the real part
of the solutions to the dispersion relation in the high frequency
limit for fundamental, surface, and body modes is given by
\begin{equation}
\frac{\omega }{k}\approx \frac{u_{j}\pm v_{wj}}{1\pm
v_{wj}u_{j}/c^{2}}~.  
\label{25}
\end{equation}
and describes sound waves $v_{wj}=a_{j}$ or Alfv\'{e}n waves $
v_{wj}=v_{Aj}$ propagating with and against the jet flow inside the jet.
Unstable growing solutions are associated with the backwards moving (in the
jet fluid reference frame) wave but the growth rate is vanishingly small in
this limit.

\vspace{-0.5cm}
\section{Numerical Solution of the Dispersion Relation}

The detailed behavior of solutions within an order of magnitude of the
resonant frequency and for comparable sound and Alfv\'{e}n wave speeds
must be investigated by numerical solution of the dispersion relation.
Analytical solutions found in the previous section can be used
for initial estimates and to provide the functional behavior
of solutions. Numerical solution of the dispersion relation also
allows a determination of the accuracy and applicability of the
analytical expressions in \S 3.

In this section pinch fundamental, helical surface and elliptical
surface, and the associated first body modes are investigated in the
fluid, magnetic and magnetosonic regimes.  These modes are chosen as
they have been identified with structure seen in relativistic
hydrodynamic ({\bf RHD}) numerical simulations or tentatively
identified with structures in resolved AGN jets. For example, trailing
shocks in a numerical simulation (Agudo et al.\ 2002) and in the
3C\,120 jet (G\'omez et al.\ 2001) have been identified with the first
pinch body mode. The development of large scale helical twisting of
jets has been attributed to or may be associated with growth of the
helical surface mode, e.g., 3C\,449 (Hardee 1981) and Cygnus A (Hardee
1996) Additionally, the development of twisted filamentary structures
has been attributed to helical and elliptical surface and first body
modes, e.g., 3C\,273 (Lobanov \& Zensus 2001), M\,87 (Lobanov, Hardee
\& Eilek 2003), 3C\,120 (Hardee, Walker \& G\'omez 2005), and have been
studied in RHD numerical simulations, e.g., Hardee \& Hughes (2003);
Perucho et al.\ (2006).

\vspace{-0.5cm}
\subsection{Fluid Limit}

In this section the basic behavior of the pinch (F) fundamental,
helical (S) surface and elliptical (S) surface modes is investigated:
(1) as a function of varying sound speed in the external sheath or jet
spine for a fixed sound speed in the jet spine or external sheath and
no sheath flow, (2) as a function of equal sound speeds in the jet
spine and external sheath for no sheath flow, and (3) as a function of
sheath flow for a relatively high sound speed equal in jet spine and
external sheath. In general only growing solutions are shown and
complexities associated with multiple crossing solutions are not
shown. For all solutions shown the jet spine Lorentz factor and speed
are set to $\gamma = 2.5$ and $u_j = 0.9165~c$.  Sound speeds are
input directly with the only constant being the sheath number density.
Total pressure and spine density are quantities computed for the
specified sound speeds. The adiabatic index is chosen to be $\Gamma =
13/9$ when $0.1 \le a_{j,e}/c \le 0.5$ consistent with relativistically
hot electrons and cold protons (Synge 1957).  For sound speeds
$a_{j,e} \sim c/\sqrt 3$ the adiabatic index is set to $\Gamma =
4/3$. Solutions shown assume zero magnetic field. Test calculations
with magnetic fields giving magnetic pressures a few percent of the
gas pressure and Alfv\'en wave speeds an order of magnitude less than the
sound speeds give almost identical results.

In Figure 1 solutions in the left column are for a fixed jet spine
sound speed $a_j = 0.3~c$ and in the right column are for a fixed
external sheath sound speed $a_e = 0.3~c$.  The solutions shown in
Figure 1 confirm the accuracy of the low frequency solutions to the
pinch fundamental mode, eqs.\ (5 \& 6), and the helical and elliptical
surface modes, eq.\ (8). Note that fast or slow wave speeds are
possible at low frequencies depending on whether $\eta \simeq
(\gamma_j a_e/\gamma_e a_j)^2$ in eq.\ (8) is much greater or much
less than one, respectively. The numerical solutions to the dispersion
relation show that the maximum growth rate is primarily a function of
the jet spine sound speed and only secondarily a function of the
external sheath sound speed as indicated by eqs.\ (18 - 20).  Where a
distinct supersonic resonance exists, the resonant frequency is
primarily a function of the external sheath sound speed as predicted
from eq.\ (16).  The analytical expression for the resonant frequency
for the helical and elliptical surface modes provides the correct
functional variation to within a constant multiplier provided $a_e \le
c/\sqrt 3$ and $a_j < c/3$. A dramatic increase in the resonant
frequency and modest increase in the growth rate for larger jet spine
sound speeds indicates the transition to transonic behavior.  Equation
(15) for the resonant wave speed and equation (17) for the resonant
wavelength also provide a reasonable approximation to the functional
variations provided $a_e \le c/\sqrt 3$ and $a_j < c/3$. These results
confirm the resonant solutions found in \S 3.2. At frequencies more
than an order of magnitude above resonance the growth rate is greatly
reduced and solutions approach the high frequency limiting form given
by eq.\ (26). Note that eq.\ (26) allows only relatively high wave
speeds at high frequencies because $a_j \le c/\sqrt 3$.

In Figure 2 the behavior of solutions to the fundamental/surface (left
column) and associated first body mode (right column) shows how
solutions change as the sound speed increases in both the jet spine
and external sheath.  Here I illustrate the transition from supersonic
to transonic behavior for no flow in the sheath.  At low frequencies
the modes behave as predicted by the analytic solutions given in \S
3.1. The solutions show the expected shift to a higher resonant
frequency that is primarily a function of the increased external
sheath sound speed and an accompanying increase in the resonant growth
rate that is primarily a function of the increased jet spine sound
speed. The resonance disappears as sound speeds approach $c/\sqrt 3$
as the jet becomes transonic as predicted by the resonance condition
in \S 3.2.  
\begin{figure}[hp!]
\vspace{18.7cm} 
\includegraphics{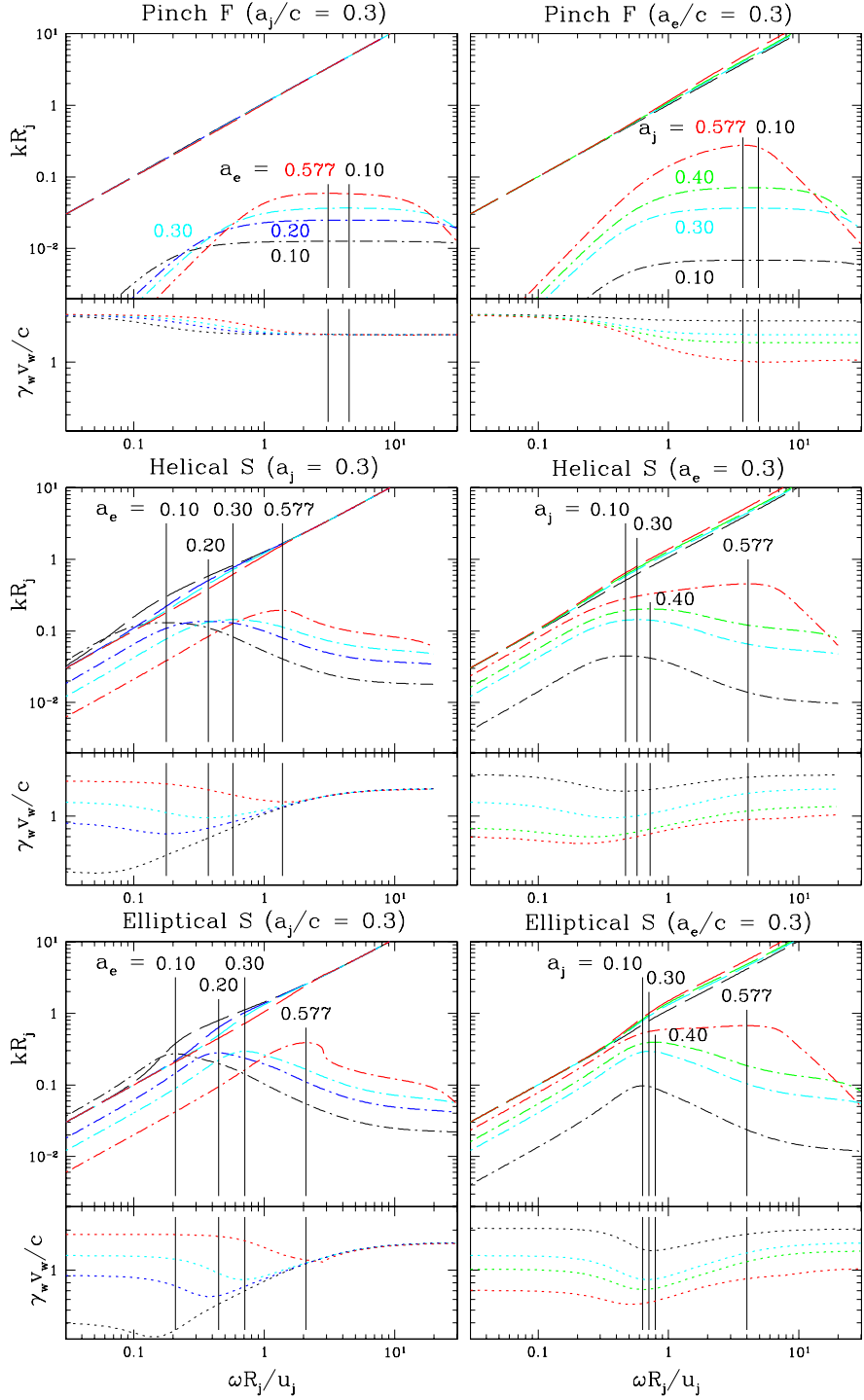} 
\caption{\footnotesize \baselineskip 11pt Solutions to the dispersion
relation for pinch fundamental, helical surface and elliptical surface
modes for different sound speeds in the sheath (left column) and in
the spine (right column) are shown for no sheath flow. The real part
of the wavenumber, $k_rR_j$, is shown by the dashed lines and the
imaginary part , $k_iR_j$, is shown by the dash-dot lines as a
function of the dimensionless angular frequency, $\omega R_j/
u_j$. For the pinch mode the vertical lines indicate the maximum
growth rate range.  Otherwise, the vertical lines indicate the
location of maximum growth. Immediately under the dispersion relation
solution panel is a panel that shows the relativistic wave speed,
$\gamma_w v_w/c$. Line colors indicate the sound speed in units of c:
(black) 0.10, (blue) 0.20, (cyan) 0.30, (green) 0.40, \& (red) 0.577.
\label {f1}} 
\end{figure}
\begin{figure}[hp!]
\vspace{18.7cm} 
\includegraphics{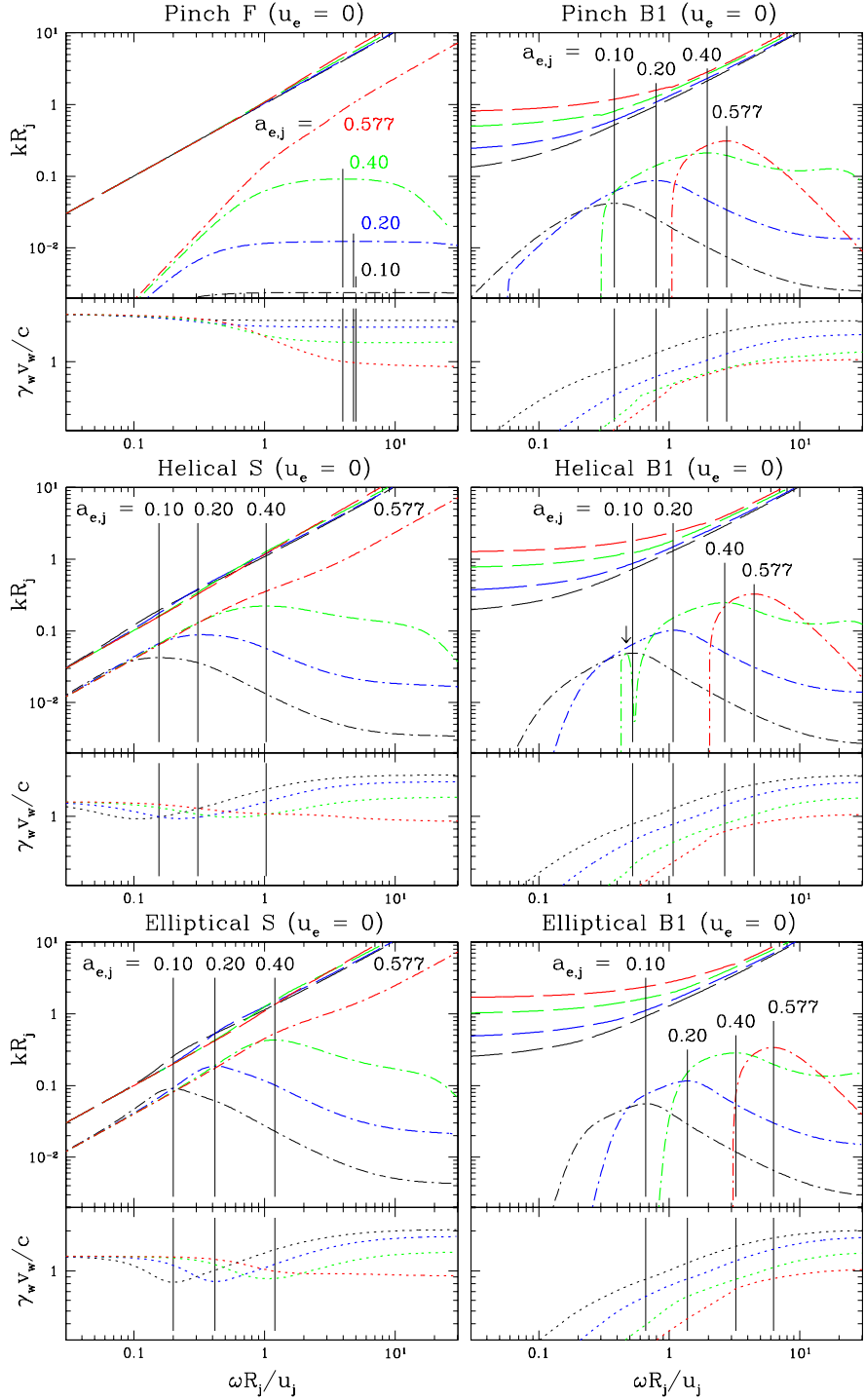} 
\caption{\footnotesize \baselineskip 11pt Solutions to the dispersion
relation for pinch fundamental, helical surface, elliptical surface
(left column) and the first body (right column) modes are shown for
equal sound speeds in spine and sheath and no sheath flow.  Real and
imaginary parts of the wavenumber as a function of angular frequency
are shown as in Figure 1.  Locations of the maximum growth rate are
indicated by the vertical solid lines. A vertical arrow (helical B1)
indicates a narrow damping feature. The underlying panel shows the
relativistic wave speed, $\gamma_w v_w/c$. Line colors indicate the
sound speed in units of c: (black) 0.10, (blue) 0.20, (green) 0.40, \&
(red) 0.577.
\label {f2}} 
\end{figure}
\begin{figure}[hp!]
\vspace{18.7cm} 
\includegraphics{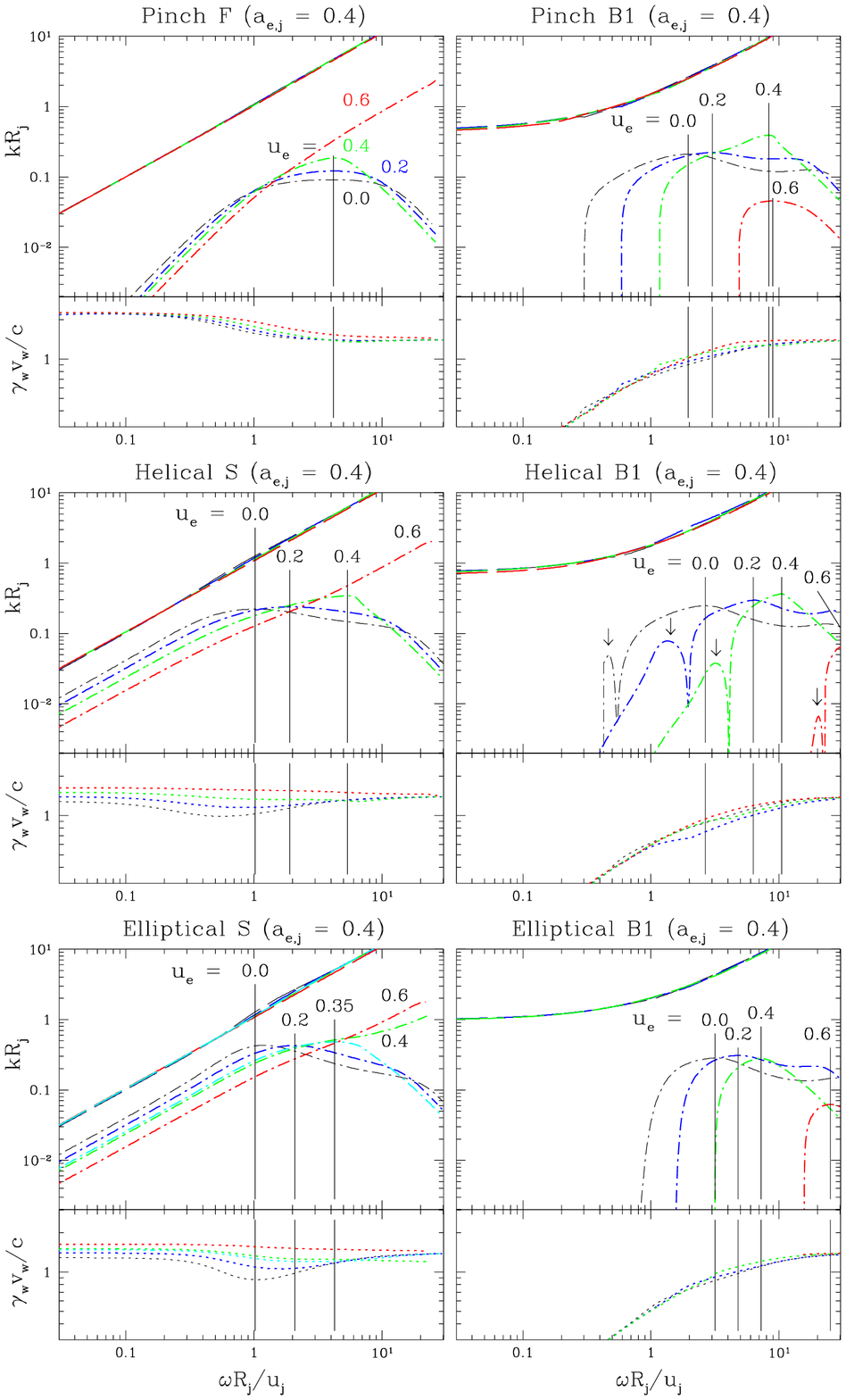} 
\caption{\footnotesize \baselineskip 11pt Solutions to the dispersion
relation for pinch fundamental, helical surface, elliptical surface
(left column) and first body (right column) modes as a function of the
sheath speed for equal sound speeds in spine and sheath. Real and
imaginary parts of the wavenumber as a function of angular frequency
are shown as in Figure 1. Locations of the maximum growth rate are
indicated by the vertical solid lines. Vertical arrows (helical B1)
indicate damping features.  The underlying panel shows the
relativistic wave speed, $\gamma_w v_w/c$. Line colors indicate the
sheath speed in units of c: (black) 0.0, (blue) 0.20, (cyan) 0.35,
(green) 0.40, \& (red) 0.60.
\label {f3}} 
\end{figure}
In the transonic regime high frequency fundamental/surface
mode growth rates and wave speeds are identical with wave speeds given
by eq.\ (26). Provided the jet is sufficiently supersonic, i.e.,
$a_{j,e} < 0.5~c$, the maximum growth rate of the first body mode is
greater than that of the pinch fundamental mode, is comparable to that
of the helical surface mode, and is less than that of the elliptical
surface mode. A narrow damping peak shown for the helical first body
(B1) solution when $a_{j,e} = 0.4~c$ is indicative of complexities in
the body mode solution structure. In the transonic regime growth of
the first body mode is less than that of the pinch fundamental,
helical surface and elliptical surface modes.

Figure 3 illustrates the behavior of fundamental/surface and first
body modes as a function of the sheath speed for equal sound speeds in
spine and sheath of $a_{j,e} = 0.4~c$.  For this value of the sound
speeds a sheath speed $u_e = 0$ provides a supersonic solution
structure baseline.  At low frequencies the surface modes behave as
predicted by eq.\ (8), and the wave speed rises as $u_e$ increases. As
$u_e$ increases the resonant frequency increases in accordance with
eq.\ (16).  On the other hand, the growth rate at resonance does not
vary significantly in accordance with eqs.\ (18 \& 19).  When the
sheath speed exceeds the sound speed, solutions make a transition from
supersonic to transonic structure.  Note that the transition point
between supersonic and transonic behavior is similar but not identical
for the helical and elliptical surface modes, i.e., ocurs at a
slightly lower sheath speed for the elliptical mode. The first body
modes also show an increase in resonant frequency with little change
in the maximum growth rate provided the sheath speed remains below the
sound speed. A significant damping feature in the helical first body
(B1) panel, is found. While a similar damping feature was not found
for the pinch and elliptical first body mode, this does not indicate a
significant difference as the root finding technique does not find all
structure associated with the body modes.  The body mode solution
structure is complex with multiple solutions not shown here and modest
damping or growth can occur where solutions cross, e.g., Mizuno,
Hardee \& Nishikawa (2006). When the sheath speed exceeds the sound
speed the maximum body mode growth rate delines significantly. This
result is quite different from the transonic solution behavior
illustrated in Figure 2 when $a_{j,e} = 0.577~c$ for no sheath flow.
Thus, sheath flow effects stability of the relativistic jet beyond
that accompanying an increase to the maximum sound speed in the
absence of sheath flow.  The reduction in growth of the body modes in
the presence of sheath flow provides the relativistic jet equivalent
of non-relativistic transonic/subsonic jet solution behavior. At the
higher frequencies wave speeds are identical, with wave speeds given
by eq.\ (26). Note that the high frequency wave speeds are nearly
independent of $u_e$.

\vspace{-0.5cm}
\subsection{Magnetic Limit}

In this subsection the basic behavior of pinch, helical and elliptical
modes is investigated: (1) as a function of varying Alfv\'en speed in
the external sheath or jet spine for a fixed Alfv\'en speed in the jet
spine or external sheath and no sheath flow, (2) as a function of
equal Alfv\'en speeds in the jet spine and external sheath for no
sheath flow, and (3) as a function of sheath speed for a relatively
high Alfv\'en speed equal in jet spine and external sheath. In general
only growing solutions are shown and complexities associated with
multiple crossing solutions are not shown. For all solutions shown the
jet spine Lorentz factor and speed are set to $\gamma = 2.5$ and $u_j
= 0.9165~c$.  Alfv\'en speeds are on the order of two magnitudes
larger than the sound speed and are determined by varying the sound
speeds but with a gas pressure fraction on the order of 0.01\% of the
total pressure. Only the sheath number density is held constant.  The
adiabatic index is set to $\Gamma = 5/3$ when $a_{j,e}/c << 0.1$
consistent with low gas pressures and temperatures.
\begin{figure}[h!]
\vspace{13.0cm} 
\includegraphics{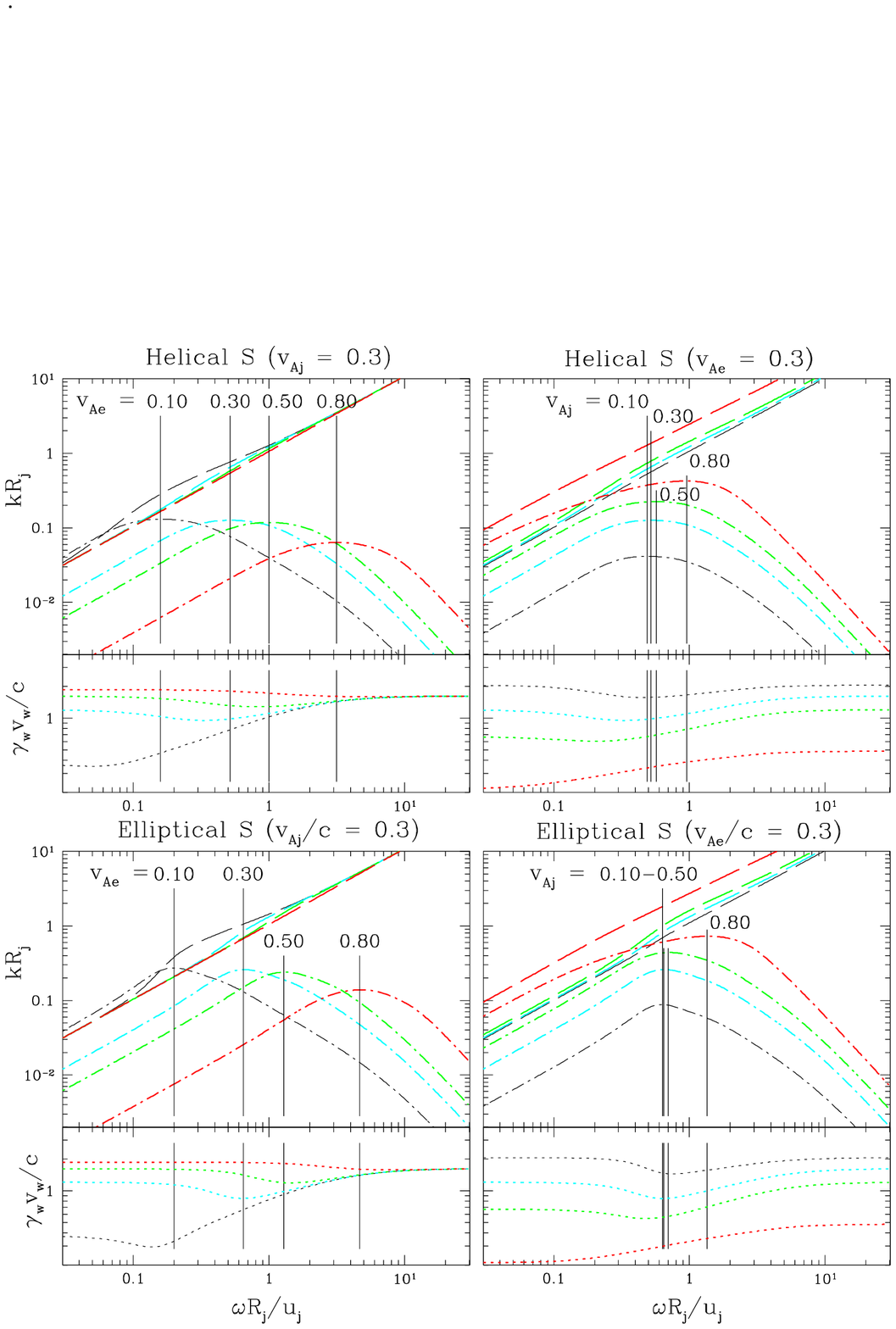} 
\caption{\footnotesize \baselineskip 11pt Solutions to the dispersion
relation for pinch fundamental, helical surface and elliptical surface
modes for different Alfv\'en speeds in the sheath (left column) and in
the spine (right column) are shown for no sheath flow. Sound speeds
are $a_{j,e} \sim 0.01 v_{Aj,e}$. As in Figures 1 - 3, the real part
of the wavenumber, $k_rR_j$, is shown by the dashed lines and the
imaginary part , $k_iR_j$, is shown by the dash-dot lines as a
function of the dimensionless angular frequency, $\omega R_j/
u_j$. The vertical lines indicate the location of maximum growth. The
underlying panel shows the relativistic wave speed, $\gamma_w
v_w/c$. Line colors indicate the Alfv\'en speed in units of c: (black)
0.10, (cyan) 0.30, (green) 0.50, \& (red) 0.80.
\label {f4}} 
\end{figure}

The solutions shown in Figure 4 confirm the theoretical predictions in
the magnetic limit with behavior depending on the Alfv\'en speed like
the behavior found for the sound speed (see Figure 1).  The pinch
fundamental mode (not shown) has a growth rate almost entirely
dependent on sound speeds and is negligable in the magnetic limit as
predicted by eq.\ (6). In Figure 4 solutions in the left column are
for a fixed jet spine Alfv\'en speed $v_{Aj} = 0.3~c$ and in the right
column are for a fixed external sheath Alfv\'en speed $v_{Ae} =
0.3~c$.  The solutions shown confirm the accuracy of the low frequency
solutions for helical and elliptical surface modes given by eq.\
(8). Note that low frequency wave speeds can be high or low depending
on the values of $\eta = \gamma_j^2 W_j/\gamma_e^2 W_e$,
$V_{Ae}/\gamma_e$ and $V_{Aj}/\gamma_j$. The numerical solutions to
the dispersion relation show that the maximum growth rate is primarily
a function of the jet spine Alfv\'en speed and only secondarily a
function of the external sheath Alfv\'en speed as predicted by eqs.\
(18 - 20). The resonant frequency is primarily a function of the
external sheath Alfv\'en speed as predicted by eq.\ (16).  The
analytical expression for the resonant frequency of the helical and
elliptical surface modes provides the correct functional variation to
within a constant multiplier provided $v_{Aj,e} < 0.5~c$.  Decrease in
the growth rate for jet sheath Alfv\'en speeds $v_{Ae} > 0.5~c$
indicates the transition towards trans-Alfv\'enic behavior.  Equation
(15) for the resonant wave speed and equation (17) for the resonant
wavelength also provide a reasonable approximation to the functional
variations for $v_{Aj,e} < 0.5~c$.  
At frequencies more than an order
of magnitude above resonance the growth rate is greatly reduced and
solutions approach the high frequency limiting form given by eq.\
(26). The surface modes have relatively slow wave speeds, $\gamma_w
v_w/c < 1$ at high frequencies when the Alfv\'en wave speed $v_{Aj}
> 0.5~c$.  Unlike the fluid case, the helical and elliptical
surface modes are stabilized for Alfv\'en speeds somewhat in excess of
$v_{Aj,e} \sim 0.8~c$ in accordance with eqs.\ (8 \& 24).

In Figure 5 the behavior of solutions to the pinch fundamental mode is
shown in addition to the helical and elliptical surface (left column)
and associated first body modes (right column) and the figure shows
how solutions change as the Alfv\'en speed increases in both the jet
spine and external sheath. The sound speed is $a_{j,e} = 0.2~c$ for
\begin{figure}[hp!]
\vspace{18.7cm} 
\includegraphics{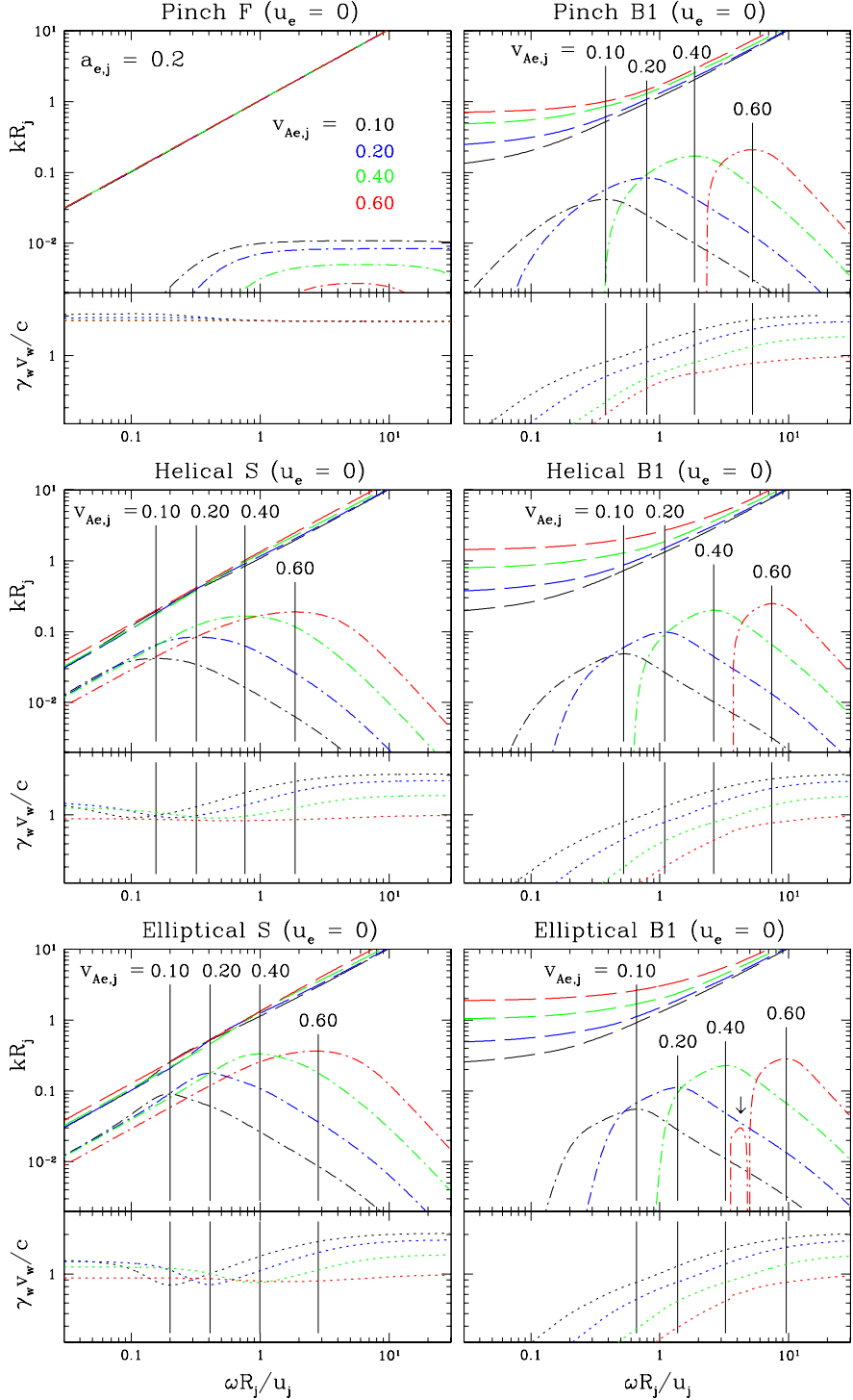} 
\caption{\footnotesize \baselineskip 11pt Solutions to the dispersion
relation for pinch fundamental, helical surface, elliptical surface
(left column) and the first body (right column) modes are shown for
equal sound speeds in jet and sheath and no sheath flow. Pinch
fundamental mode sound speed is $a_{j,e} = 0.2~c$. Sound speeds for
all other cases are $a_{j,e} \sim 0.01 v_{Aj,e}$. Real and imaginary
parts of the wavenumber as a function of angular frequency are shown
as in Figure 4.  Locations of the maximum growth rate are indicated by
the vertical solid lines. A vertical arrow (elliptical B1) indicates a
narrow damping feature. The underlying panel shows the relativistic
wave speed, $\gamma_w v_w/c$. Line colors indicate the Alfv\'en speed
in units of c: (black) 0.10, (blue) 0.20, (green) 0.40, \& (red) 0.60.
\label {f5}} 
\end{figure}
\begin{figure}[hp!]
\vspace{18.7cm} 
\includegraphics{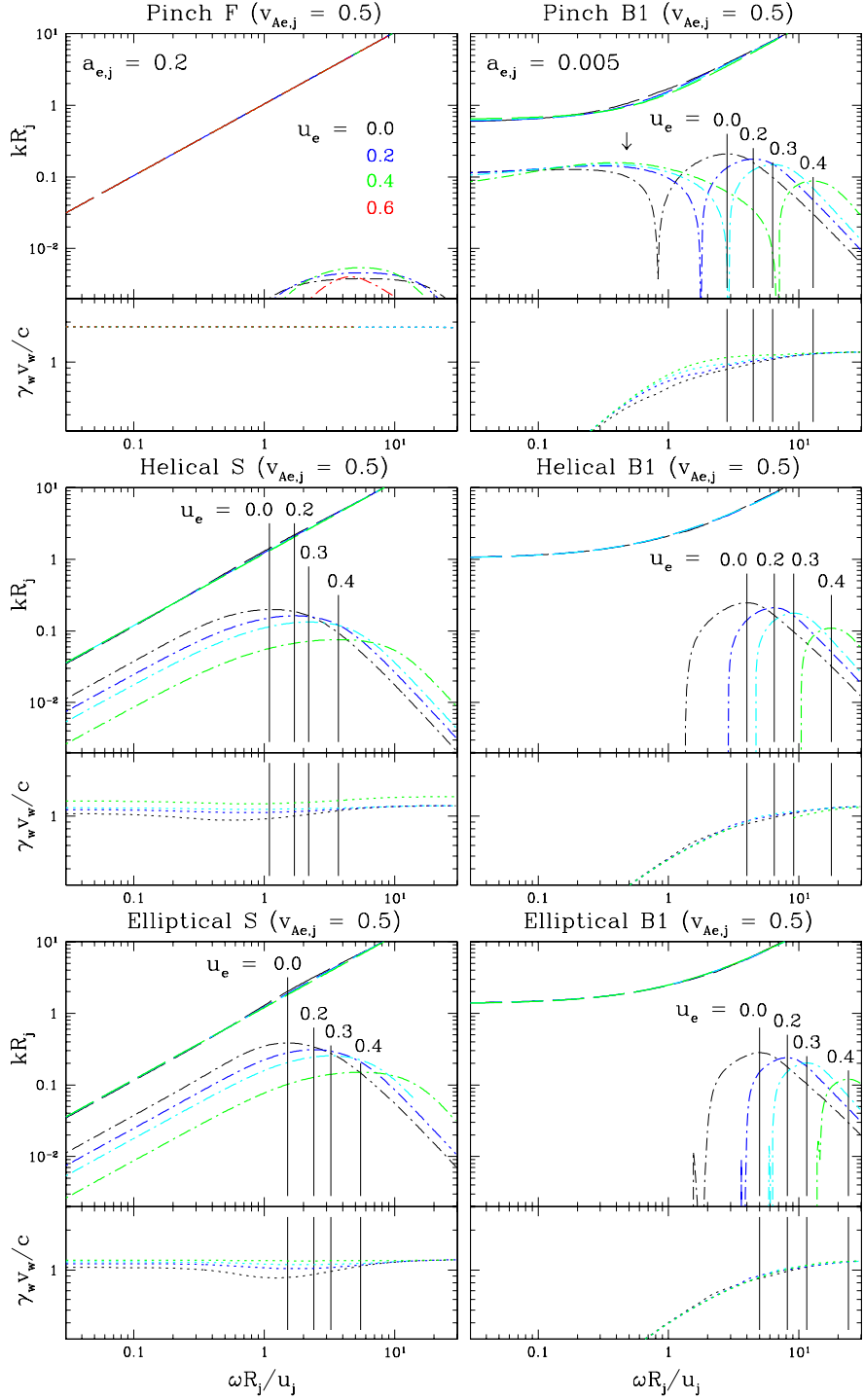} 
\caption{\footnotesize \baselineskip 11pt Solutions to the dispersion
relation for pinch fundamental, helical surface, elliptical surface
(left column) and the first body (right column) modes are shown for
equal sound speeds in jet and sheath for different sheath flow
speeds. The pinch fundamental mode sound speed is $a_{j,e} =
0.2~c$. Sound speeds for all other cases are $a_{j,e} \sim 0.01
v_{Aj,e} = 0.005~c$. Real and imaginary parts of the wavenumber as a
function of angular frequency are shown as in Figure 4.  Locations of
the maximum growth rate are indicated by the vertical solid lines. A
vertical arrow indicates low frequency damping of the pinch B1
solutions. The underlying panel shows the relativistic wave speed,
$\gamma_w v_w/c$. Line colors indicate the sheath speed in units of c:
(black) 0.0, (blue) 0.20, (cyan) 0.30, (green) 0.40, \& (red) 0.60.
\label {f6}} 
\end{figure}
the pinch fundamental mode panel in order to illustrate the mode
behavior with increasing Alfv\'en speed.  Sound speeds for all body
modes and for helical and elliptical surface modes are $a_{j,e} \sim
0.01 v_{Aj,e}$. Here the transition from super-Alfv\'enic towards
trans-Alfv\'enic behavior for no flow in the sheath is illustrated.
At low frequencies the modes behave as predicted by the analytic
solutions given in \S 3.1. The growth rate of the pinch fundamental
mode is reduced as the Alfv\'en speed increases as predicted by eq.\
(6). The surface and body mode solutions show the expected shift to a
higher resonant frequency that is primarily a function of the
increased sheath Alfv\'en speed and an accompanying increase in the
resonant growth rate that is primarily a function of the increased
spine Alfv\'en speed. The resonance moves to higher frequency but the
maximum growth rate is reduced for Alfv\'en speeds $v_{Aj,e} > 0.60~c$
and all modes become stable at higher Alfv\'en speeds in accordance
with eqs.\ (8 \& 24).  At high frequencies wave speeds are given by
eq.\ (26). Provided the jet is sufficiently super-Alfv\'enic, i.e.,
$v_{Aj,e} < 0.6~c$, the maximum growth rate of the first body mode is
much greater than that of the pinch fundamental mode, is comparable to
that of the helical surface mode, and is less than that of the
elliptical surface mode. A narrow damping peak shown for the
elliptical body mode (B1) solution when $v_{Aj,e} = 0.6~c$ indicated
by the arrow is indicative of complexities in the body mode solution
structure.

Figure 6 illustrates the behavior of fundamental/surface and first
body modes as a function of the sheath speed for an equal Alfv\'en
speed in spine and sheath of $v_{Aj,e} = 0.5~c$.  For this value of
the Alfv\'en speeds a sheath speed $u_e = 0$ provides a
super-Alfv\'enic solution structure baseline. The sound speed is
$a_{j,e} = 0.2~c$ for the pinch fundamental mode panel in order to
illustrate the mode behavior with increasing sheath speed.  Sound
speeds for all other cases are $a_{j,e} \sim 0.01 v_{Aj,e}$.
Solutions in the first pinch body mode panel show the damping solution
as opposed to the purely real solution at the lower frequencies
(indicated by the arrow).  At higher frequencies the body mode is
growing. At low frequencies the surface modes behave as predicted by
the analytic solutions given in \S 3.1 and the growth rate of the
surface modes decreases as $u_e$ increases.  Additionally, the growth
rate at resonance decreases as expected for this relatively high
Alfv\'en speed as the sheath speed increases.  At the higher
frequencies wave speeds are identical, with wave speeds given by eq.\
(26). Note that the high frequency wave speeds are relatively
independent of $u_e$. When the velocity shear speed drops to less than
the ``surface'' Alfv\'en speed, see eq.\ (11b), the helical and
elliptical surface modes and the first body modes are stabilized. This
surface and body mode mode stabilization occurs when sheath speeds
exceed $u_e \sim 0.5~c$.  However, note that the maximum pinch
fundmental mode growth rate is insensitive to the sheath speed and
remains unstable at $u_e = 0.6~c$ even when all other modes are
stabilized.

\vspace{-0.5cm}
\subsection{A High Sound and Alfv\'en Speed Magnetosonic Case}

In this subsection the basic behavior of the pinch fundamental,
helical surface, elliptical surface and associated first body modes is
illustrated for different sheath speeds. 
\begin{figure}[hp!]
\vspace{18.7cm} 
\includegraphics{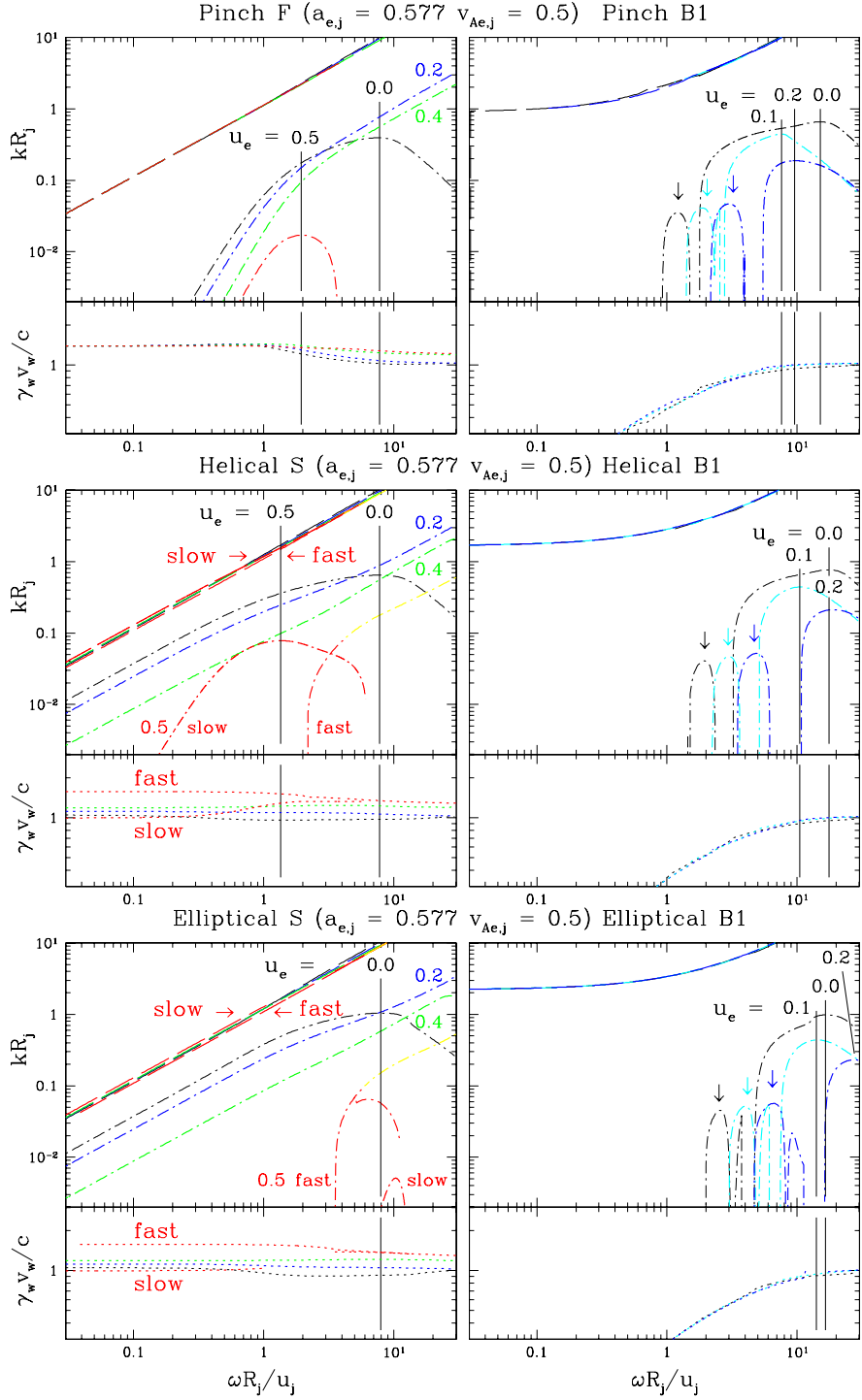} 
\caption{\footnotesize \baselineskip 11pt Solutions to the dispersion
relation for pinch fundamental, helical surface, elliptical surface
(left column), and the associated first body (right column) modes are
shown for a maximal spine and sheath sound speed, $a_{j,e} = 0.577~c$,
and a slightly smaller spine and sheath Alfv\'en speed, $v_{Aj,e} =
0.5~c$, for different sheath flow speeds. As in previous figures the
real part of the wavenumber, $k_rR_j$, is shown by the dashed lines,
the imaginary part , $k_iR_j$, is shown by the dash-dot lines, and the
vertical lines indicate the location of maximum growth. Arrows
indicate damping features. The underlying panel shows the relativistic
wave speed, $\gamma_w v_w/c$. Line colors indicate the sheath speed in
units of c: (black) 0.0, (blue) 0.20, (green) 0.40, \& (red)
0.50. Fast and slow refer to the faster and slower moving solutions
and the yellow extension indicates a damped solution.
\label {f7}} 
\end{figure}
The sheath speeds span a
solution structure from supersonic to transonic but still
super-Alfv\'enic flow.  Here the sound speed in jet spine and external
sheath are set equal with $a_{j,e} = 0.577~c$ and Alfv\'en speeds are
set equal with $v_{Aj,e} = 0.5~c$.  The solutions for this case are
shown in Figure 7.  With no sheath flow the fundamental/surface and
first body modes show a typical supersonic and super-Alfv\'enic
structure albeit the pinch fundamental mode now has a maximum growth
rate comparable to the helical and elliptical surface modes as a
consequence of the high sound speed. The associated first body modes
also have maximum growth rates comparable to the fundamental/surface
modes. Increase in the sheath speed results in a decrease in the
growth rate of the helical and elliptical surface modes at low
frequencies as predicted
by eq.\ (8).  The low frequency growth rate of the pinch fundamental
also declines with increasing sheath speed.  The resonant frequency
increases with increasing sheath speed as expected from the analytical
and numerical studies performed in the fluid and magnetic limits and
the fundamental/surface modes take on a transonic structure for sheath
speeds $0.4~c \le u_e \le 0.1~c$.

At high frequencies the fundamental/surface modes exhibit very high
growth rates provided sheath flow remains below the Alfv\'en speed. On
the other hand, the maximum growth rate of the first body modes
declines as the sheath speed increases and is reduced severely when
$u_e > 0.1~c$. This behavior is similar to what is found for
non-relativistic jets as flow enters the transonic and
super-Alfv\'enic regime (Hardee \& Rosen 1999).  Additional increase in the
sheath flow speed to $u_e > 0.4~c$ results in a decrease in the growth
rate of the fundamental/surface modes.  Solutions for the helical and
elliptical surface modes shown in Figure 7 for a sheath speed $u_e =
0.5~c$ equal to the Alfv\'en speed illustrate some of the complexity
associated with barely super-Alfv\'enic flow.  Here limited growth is
associated with both the slow and fast helical and elliptical surface
solution pair. At slower sheath speeds in the super-Alfv\'enic regime
growth is associated with the slow surface solution, i.e., backwards
moving in the jet fluid reference frame. The yellow dash-dot line
extension at higher frequencies in the helical and elliptical surface
panels indicates a damped solution. Solutions were very difficult to
follow in this parameter regime and it is possible that some solutions
were not found. When the sheath speed $u_e > 0.5~c$ all modes are
stabilized.

A choice of Alfv\'en speeds greater than sound speeds results in a
more magnetic like solution structure like that shown in \S 4.2.
A choice of Alfv\'en speeds more than a factor of two less than sound
speeds produces a more fluid like solution structure like that shown
in \S 4.1.  The more complicated solution structure illustrated
in Figure 7 only occurs for a relatively narrow range of high sound
speeds with similar or slightly lesser Alfv\'en speeds.  In general,
the detailed solution structure for situations in which sound and
Alfv\'en speeds are comparable must by examined individually, e.g.,
Mizuno, Hardee \& Nishikawa (2006), and further investigation of these
cases is beyond the scope of the present paper.

\vspace{-0.5cm}
\section{Summary}

The analytical and numerical work performed here provides for the
first time a detailed analysis of the KH stability properities of a
RMHD jet spine-sheath configuration that allows for relativistic
motions of the sheath, sound speeds up to $c/\sqrt 3$, and, by keeping
the displacement current in the analysis, Alfv\'en wave speeds
approaching lightspeed and large Alfv\'en Lorentz factors.  In the
fluid limit, the present results confirm an earlier more restricted
low frequency analytical and numerical simulation study performed by
Hardee \& Hughes (2003).  Provided the jet spine is super-sonic and
super-Alfv\'enic internally and also relative to the sheath, the
helical, elliptical and higher order surface modes and the pinch,
helical, elliptical and higher order first body modes have a maximum
growth rate at a resonant frequency.  The pinch fundamental growth
rate is significant only when the sound speeds, $a_{j,e} \sim c/\sqrt
3$. In general, the first body mode maximum growth rate is: greater
than the pinch fundamental mode, slightly greater than the helical
surface mode, slightly less than the elliptical surface mode, and
occurs at a higher frequency than the maximum growth rate for the
fundamental/surface mode.

The basic KH stability behavior as a function of spine-sheath
parameters is indicated by the analytic low frequency surface mode
solution and by the behavior of the resonant frequency.  The analytic
surface mode solution valid at frequencies below resonance is given by
\begin{equation}
\frac{\omega }{k}= \frac{\omega_{r} }{k} \pm i\frac{\omega_{i} }{k} = \frac{\left[ \eta u_{j}+u_{e}\right] \pm i\eta
^{1/2}\left[ \left( u_{j}-u_{e}\right) ^{2}-V_{As}^{2}/\gamma
_{j}^{2}\gamma _{e}^{2} \right] ^{1/2}}{(1+V_{Ae}^{2}/\gamma
_{e}^{2}c^{2})+\eta (1+V_{Aj}^{2}/\gamma _{j}^{2}c^{2})} 
\label{27}
\end{equation}
where
\begin{equation}
V_{As}^{2}\equiv \left( \gamma _{Aj}^{2}W_{j}+\gamma
_{Ae}^{2}W_{e}\right) \frac{B_{j}^{2}+B_{e}^{2}}{4\pi W_{j}W_{e}}~,
\label{28}
\end{equation}
and $\eta \equiv \gamma _{j}^{2}W_{j}\left/ \gamma
_{e}^{2}W_{e}\right.$, $V_{A}^2 \equiv B^2/4 \pi W$, $W\equiv \rho
+\left[ \Gamma /\left( \Gamma -1\right) \right] P/c^{2}$ and $\gamma
_{A}\equiv (1-v_{A}^{2}/c^{2})^{-1/2}$. Equation (27) provides a
temporal growth rate, $\omega_{i}(k)$, and a wave speed, $v_{w} =
\omega_{r}/k$. The reciprocal provides a spatial growth rate
$k_{i}(\omega)$, and growth length $\ell = k_{i}^{-1}$. Increase or
decrease of the growth rate, dependence on physical parameters and
stabilization at frequencies/wavenumbers below resonance is directly
revealed by $\omega_{i}$ in eq.\ (27). Note that higher jet Lorentz factors reduce $\omega_{i}$ through the dependence on $\eta$.

The resonant frequency is
\begin{equation}
\omega^{\ast} \propto 
\frac{v_{we} }{ \left[\left(1 - u_{e}/v^{\ast}_w\right)^2 -
\left(v_{we}/v^{\ast}_w - u_e v_{we}/c^2\right)^2\right]^{1/2}}~,
\label{29}
\end{equation}
where $v_{w}^{\ast}$ is the wave speed at resonance, eq.\ (15).  The
resonant frequency increases as the sheath sound or Alfv\'en wave
speed, $v_{we} \equiv (a_{e} , v_{Ae})$ increases and
$\omega^{\ast}\longrightarrow \infty$ when the denominator decreases
to zero as
$$
\frac{u_{j}-u_{e}}{1-u_{j}u_{e}/c^{2}}\longrightarrow \frac{v_{wj}+v_{we}}{
1+v_{wj}v_{we}/c^{2}}~,
$$
where $v_{wj,e} \equiv (a_{j,e}, v_{Aj,e})$ in the fluid and magnetic
limits, respectively. Since eq.\ (27) applies below resonance the
overall behavior of the growth rate is indicated by $\omega_{i}$.
Thus, growth rates decline to zero as $(u_{j} - u_{e})^{2} -
V_{As}^{2}/\gamma _{j}^{2}\gamma _{e}^{2} \longrightarrow 0$. The
numerical analysis of the dispersion relation shows that the pinch
fundamental and all first body modes are comparably or more readily
stabilized and thus the jet is KH stable when
\begin{equation}
\left( u_{j}-u_{e}\right)^{2}-V_{As}^{2}/\gamma _{j}^{2}\gamma
_{e}^{2} < 0~.
\label{30}
\end{equation}
This stability condition takes on a particularly
simple form when conditions in spine and sheath are equal, i.e., 
$B_e = B_j$, $W_e = W_j$, so that $v_{A,j} = v_{A,e}$, and
with $\gamma_A \equiv \gamma_{A,e} = \gamma_{A,j}$
\begin{equation}
\gamma _{j}^{2} \gamma _{e}^{2} \left( u_{j}-u_{e}\right)^{2} < 4
\gamma_A^2 \left(\gamma_A^2 - 1\right)c^2
\label{31}
\end{equation}
indicates stability. This result implies that a trans-Alfv\'enic
relativistic jet with $\gamma_ju_j \gtrsim \gamma_Av_A$ will be KH
stable, and that even a super-Alfv\'enic jet with $\gamma_j >>
\gamma_A $ can be KH stable.

\vspace{-0.5cm}
\section{Discussion}

Formally, the present results and expressions apply only to magnetic
fields parallel to an axial spine-sheath flow in which conditions
within the spine and within the sheath are independent of radius and
the sheath extends to infinity.  A rapid decline in perturbation
amplitudes in the sheath as a function of radius, governed by the Hankel
function in the dispersion relation, suggests that the present results
will apply to sheaths more than about three times the spine radius in
thickness.

The relativistic jet is transonic in the absence of sheath flow only
for spine and sheath sound speeds $\sim c/\sqrt 3$.  Only in this
regime does the pinch fundamental have a significant growth rate and,
in general, we do not expect the pinch fundamental to grow
significantly on relativistic jets.  On the other hand, the pinch
first body mode can have a significant maximum growth rate and would
dominate any axisymmetric structure.  The elliptical and higher order
surface modes have increasingly larger maximum growth rates at
resonant frequencies higher than the helical surface mode, and the
maximum first body mode growth rates for helical and elliptical modes
are comparable to that of the surface modes.  Nevertheless, we expect
the helical surface mode to achieve the largest amplitudes in the
non-linear limit as a result of the reduced saturation amplitudes that
accompany the higher resonant frequency and shorter resonant
wavelengths associated with the higher order surface modes and all
body modes.

In astrophysical jets we expect a toroidal magnetic field component, and
possibly an ordered helical structure and accompanying flow
helicity. Jet rotation (e.g., Bodo et al.\ 1996), or a radial
velocity profile (e.g., Birkinshaw 1991) will modify the present
results but will not stabilize the helical mode.  Two dimensional
non-relativistic slab jet theoretical results, indicate that KH
stabilization occurs when the velocity shear projected on the
wavevector is less than the projected Alfv\'en speed (Hardee et al.\
1992). In the work presented here magnetic and flow field are parallel
and project equally on the wavevector which for the helical (n = 1)
and elliptical (n = 2) mode lies at an angle $\theta = tan^{-1}(n/kR)$
relative to the jet axis.  Provided magnetic and flow helicity and
radial gradients in jet spine/sheath properties are not too large we
expect the present results to remain valid where $u_{j,e}$ and
$B_{j,e}$ refer to the poloidal velocity and field components.

KH driven normal mode structures move at less than the jet speed. The
fundamental pinch mode moves backwards in the jet frame at about the
sound speed nearly independent of the sheath properites and thus moves
at nearly the jet speed in the source/observer frame.  Low frequency
and long wavelength helical and higer order surface modes are advected
with wave speed indicated by eq.\ (27) and move slowly in the
source/observer frame for light, i.e.\ $\eta \equiv \gamma
_{j}^{2}W_{j}\left/ \gamma _{e}^{2}W_{e}\right. < 1$, and/or for
magnetically dominated flows. Higher frequency (above resonance) and
shorter wavelength normal mode structures move backwards in the jet
frame at the sound/Alfv\'en wave speed, have a wave speed nearly
independent of the sheath properties, and can move slowly in the
source/observer frame only for magnetically dominated flows.

Where flow and magentic fields are parallel, current driven ({\bf CD})
modes are stable (Isotomin \& Pariev 1994, 1996). Where magnetic and
flow fields are helical CD modes can be unstable (Lyubarskii 1999) in
addition to the KH modes.  CD and KH instability are expected to
produce helically twisted structure. However, the conditions for
instability, the radial structure, the growth rate and the pattern
motions are different.  For example, KH modes grow more rapidly when
the magnetic field is force-free (e.g., Appl 1996), and
non-relativistic simulation work (e.g., Lery et al.\ 2000; Baty \&
Keppens 2003; Nakamura \& Meier 2004) indicates that CD driven
structure is internal to any spine-sheath interface and moves at the
jet speed.

The differences between KH and CD instability can serve to identify
the source of helical structure on relativistic jets and allow
determination of jet properties near to the central engine. Perhaps
the observation of relatively low proper motions in the TeV BL Lacs
when intensity modeling requires high flow Lorentz factors (Ghisellini
et al.\ 2005) is an indication of a magnetically dominated KH unstable
spine-sheath configuration.

\acknowledgments

The author acknowledges partial support through National Space Science
and Technology Center (NSSTC/NASA) cooperative agreement NCC8-256 and
by National Science Foundation (NSF) award AST-0506666 to the
University of Alabama.

\appendix 
\baselineskip 12pt
\parskip 2pt
\vspace{-0.5cm}
\section{Linearization of the RMHD Equations}

In vector notation the relativistic MHD continuity equation, energy
equation, and momentum equation can be written as:
\begin{equation}
\frac{\partial }{\partial t}\left[ \gamma \rho \right] +\mathbf{
\nabla }\cdot \left[ \gamma \rho \mathbf{v}\right] =0~,
\label{A1}
\end{equation}
\begin{equation}
\frac{\partial }{\partial t}\left[ \gamma ^{2}W-\frac{P}{c^{2}}+\frac{B^{2}}{
8\pi c^{2}}(1+\frac{v^{2}}{c^{2}})-\frac{(\mathbf{v}/c\cdot \mathbf{B})^{2}}{
8\pi c^{2}}\right] +\mathbf{\nabla }\cdot \left[ \gamma ^{2}W
\mathbf{v}+\frac{B^{2}}{4\pi c^{2}}\mathbf{v-}(\mathbf{v}\cdot \mathbf{B})
\frac{\mathbf{B}}{4\pi c^{2}}\right] =0~, \label{A2}
\end{equation}
and
\begin{equation}
\gamma ^{2}W\left( \frac{\partial }{\partial t}\mathbf{v}+\mathbf{v}\cdot
\nabla \mathbf{v}\right) =-\mathbf{\nabla }P-\frac{\mathbf{v}}{
c^{2}}\frac{\partial }{\partial t}P+\rho _{q}\mathbf{E}+\frac{\left[ \mathbf{
j}\times \mathbf{B}\right] }{4\pi }~.  \label{A3}
\end{equation}
These equations along with Maxwell's equations
\doublespace
$$
\begin{array}{cc}
\mathbf{\nabla }\cdot \mathbf{B}=0 & 
\mathbf{\nabla }\cdot \mathbf{E}=4\pi \rho _{q} \\ 
\mathbf{\nabla }\times \mathbf{B}=\frac{1}{c}\frac{\partial }{\partial t}
\mathbf{E}+\frac{4\pi }{c}\mathbf{j} & 
\mathbf{\nabla }\times \mathbf{E}=-\frac{1}{c}\frac{\partial }{\partial t}
\mathbf{B}
\end{array}
$$
\baselineskip 12pt
and assuming ideal MHD with comoving electric field equal to zero 
$$
\mathbf{E=-}\frac{\mathbf{v}\times \mathbf{B}}{c}~,
$$
provide the complete set of ideal RMHD equations.\ In the above $W$ is the
enthalpy, the Lorentz factor $\gamma =(1-\mathbf{v\cdot
  v}/c^{2})^{-1/2}$, and $
\rho $ is the proper density. In what follows I will assume that the
effects of
radiation can be ignored, the enthalpy is given by
$$
W=\rho +\frac{\Gamma }{\Gamma -1}\frac{P}{c^{2}}~,
$$
and the condition for isentropic flow is given by 
$$
\left( \frac{\partial }{\partial t}+\mathbf{v\cdot \nabla }\right)
\left( \frac{P}{\rho ^{\Gamma }}\right) =0 ~.
$$

The general approach to analyzing the time dependent properties of this
system is to linearize the ideal RMHD equations, where the density,
velocity, pressure and magnetic field are written as $\rho =\rho _{0}+\rho
_{1}$, $\mathbf{v}=\mathbf{u}+\mathbf{v}_{1}$ (we use $\mathbf{v}_{0}\equiv 
\mathbf{u}$ for notational reasons), $P=P_{0}+P_{1}$ $\mathbf{E}=\mathbf{E}
_{0}+\mathbf{E}_{1}$, and $\mathbf{B}=\mathbf{B}_{0}+\mathbf{B}_{1}$, where
subscript 1 refers to a perturbation to the equilibrium quantity with
subscript 0. Additionally, $W=W_{0}+W_{1}$, $\gamma ^{2}=(\gamma _{0}+\gamma
_{1})^{2}\simeq \gamma _{0}^{2}+2\gamma _{0}^{4}\mathbf{u\cdot v}_{1}/c^{2}$
and $\gamma _{1} \simeq \gamma _{0}^{3}\mathbf{u\cdot v}_{1}/c^{2}$. \ It is
assumed that the initial equilibrium system satisfies the zero order
equations. The linearized continuity, energy and momentum equation
become
\begin{equation}
\frac{\partial }{\partial t}\left[ \gamma _{0}\rho _{1}+\gamma _{1}\rho _{0}
\right] +\mathbf{\nabla }\cdot \left[ \gamma _{0}\rho _{1}\mathbf{u
}+\gamma _{0}\rho _{0}\mathbf{v}_{1}+\gamma _{1}\rho
_{0}\mathbf{u}\right] =0 ~,
  \label{A4}
\end{equation}
\doublespace
\begin{equation}
\begin{array}{c}
\nonumber \frac{\partial }{\partial t}\left[ \gamma _{0}^{2}W_{1}-P_{1}/c^{2}+2\gamma
_{0}^{4}\left( \mathbf{u}\cdot \mathbf{v}_{1}/c^{2}\right) W_{0}\right] +
\mathbf{\nabla }\cdot \left[ \gamma _{0}^{2}W_{1}\mathbf{u}
+2\gamma _{0}^{4}\left( \mathbf{u}\cdot \mathbf{v}_{1}/c^{2}\right) W_{0}
\mathbf{u}+\gamma _{0}^{2}W_{0}\mathbf{v}_{1}\right] \\ 
+\frac{1}{4\pi c^{2}}\frac{\partial }{\partial t}\left[ B_{0}^{2}\left( 
\mathbf{u}\cdot \mathbf{v}_{1}/c^{2}\right) +(1+u^{2}/c^{2})\mathbf{B}_{0}
\mathbf{\cdot B}_{1}-(\mathbf{u\cdot B}_{1}/c+\mathbf{v}_{1}\mathbf{\cdot B}
_{0}/c)\mathbf{u\cdot B}_{0}/c\right] \\ \nonumber
+\frac{1}{4\pi c^{2}}\mathbf{\bigtriangledown }\cdot \left[ 2(\mathbf{B}_{0}
\mathbf{\cdot B}_{1})\mathbf{u+}B_{0}^{2}\mathbf{v}_{1}-(\mathbf{u\cdot B}
_{0})\mathbf{B}_{1}-(\mathbf{u\cdot B}_{1})\mathbf{B}_{0}-(\mathbf{v}_{1}
\mathbf{\cdot B}_{0})\mathbf{B}_{0}\right] =0~,
\label{A5}
\end{array}
\end{equation}
\baselineskip 12pt
and
\begin{equation}
\gamma _{0}^{2}W_{0}\left( \frac{\partial \mathbf{v}_{1}}{\partial t}+
\mathbf{u}\cdot \nabla \mathbf{v}_{1}\right) =-\mathbf{\nabla }
P_{1}-\frac{\mathbf{u}}{c^{2}}\frac{\partial P_{1}}{\partial t}+\frac{(
\mathbf{j}_{0}\mathbf{\times B}_{1})+(\mathbf{j}_{1}\mathbf{\times B}_{0})}{c
}~.  \label{A6}
\end{equation}
The linearized Maxwell equations become:
\doublespace
$$
\begin{array}{cc}
\mathbf{\nabla }\cdot \mathbf{B}_{1} =0 & 
\mathbf{\nabla }\cdot \mathbf{E}_{1} =4\pi \rho _{q1} \\ 
\mathbf{\nabla }\times \mathbf{B}_{1} =\frac{1}{c}\frac{\partial }{\partial t}
\mathbf{E}_{1}+\frac{4\pi }{c}\mathbf{j}_{1} & 
\mathbf{\nabla }\times \mathbf{E}_{1} =-\frac{1}{c}\frac{\partial }{\partial t
}\mathbf{B}_{1}
\end{array}
$$
\baselineskip 12pt
where I keep the displacement current in order to allow for strong magnetic
fields and Alfv\'{e}n wave speeds comparable to lightspeed. Under the
assumption of ideal MHD, the comoving electric field is zero, the
equilibrium charge density $\rho _{q,0}=0$, and the electric field 
$$
\mathbf{E}_{1}\mathbf{=-}\frac{\mathbf{u}\times \mathbf{B}_{1}+\mathbf{v}_{1}
\mathbf{\times B}_{0}}{c}~,
$$
is first order, the charge density $\rho _{q1}=\left( \mathbf{
\bigtriangledown }\cdot \mathbf{E}_{1}\right) \mathbf{/}4\pi $ is also first
order, and the electrostatic force term, $\rho _{q1}\mathbf{E}_{1}$, is
second order and dropped from the linearized momentum equation. The
condition for isentropic perturbations becomes
$$
P_{1}=\tilde{a}^{2}\rho _{1}=\left( \Gamma \frac{P_{0}}{\rho _{0}}\right)
\rho _{1} ~.
$$
This basic set of linearized RMHD equations is similar to those found
in Begelman (1998) but allows a relativistic zeroth order velocity,
i.e., $\mathbf{v} = \mathbf{u} + \mathbf{v_1}$ and $\mathbf{u}
\lesssim c$ whereas Begelman allowed only for relativistic first order
motions, $\mathbf{v_1}$.

In what follows let us model a jet as a cylinder of radius R, having a
uniform proper density, $\rho _{j}$, a uniform axial magnetic field, $
B_{j}=B_{z,j}$, and a uniform velocity, $u_{j}=u_{z,j}$. The external medium
is assumed to have a uniform proper density, $\rho _{e}$, a uniform axial
magnetic field, $B_{e}=B_{z,e}$, and a uniform velocity, $u_{e}=u_{z,e}$. An
external velocity could be the result of a wind or sheath outflow around a central
jet, $u_{e}>0$, or could represent backflow, $u_{e}<0$, in a cocoon
surrounding the jet. The jet is established in static total pressure balance
with the external medium where the total static uniform pressure is $
P_{e}^{\ast }\equiv P_{e}+B_{e}^{2}/8\pi =P_{j}^{\ast }\equiv
P_{j}+B_{j}^{2}/8\pi $. \ Under these assumptions the linearized continuity
equation becomes
\begin{equation}
\frac{\partial }{\partial t}\left[ \gamma _{0}\rho _{1}+\gamma _{1}\rho _{0}
\right] +u\frac{\partial }{\partial z}\left[ \gamma _{0}\rho _{1}+\gamma
_{1}\rho _{0}\right] +\gamma _{0}\rho _{0}\mathbf{\nabla }\cdot 
\mathbf{v}_{1}=0~.  \label{A7}
\end{equation}
The linearized energy equation becomes
\begin{equation}
\frac{\partial }{\partial t}\left[ \gamma _{0}^{2}W_{1}-\frac{P_{1}}{c^{2}}
+2\gamma _{0}^{4}\frac{uv_{z1}}{c^{2}}W_{0}\right] +u\frac{\partial }{
\partial z}\left[ \gamma _{0}^{2}W_{1}+2\gamma _{0}^{4}\frac{uv_{z1}}{c^{2}}
W_{0}\right] +\gamma _{0}^{2}W_{0}\mathbf{\nabla }\cdot \mathbf{v}
_{1}=0~.  \label{A8}
\end{equation}
This result for the linearized energy equation is found by noting that the
magnetic terms in the energy equation linearize to
\doublespace
$$
\begin{array}{c}
B_{0}\left[ \frac{\partial }{\partial t}B_{z1}+u\frac{\partial }{\partial z}
B_{z1}\right] -(uB_{0})\mathbf{\nabla }\cdot \mathbf{B}
_{1}+B_{0}^{2}\mathbf{\nabla }\cdot \mathbf{v}_{1}-B_{0}^{2}\frac{
\partial }{\partial z}v_{z1}= \\ 
-B_{0}^{2}\left[ \mathbf{\nabla }\cdot \mathbf{v}_{1}-\frac{
\partial }{\partial z}v_{z1}\right] +B_{0}^{2}\left[ \mathbf{
\nabla }\cdot \mathbf{v}_{1}-\frac{\partial }{\partial z}v_{z1}
\right] =0
\end{array}
$$
\baselineskip 12pt
where I have used
$$
\frac{\partial }{\partial t}B_{z1}+u\frac{\partial }{\partial z}B_{z1}=-
\frac{B_{0}}{r}\left[ \frac{\partial }{\partial r}(rv_{r1})+\frac{\partial }{
\partial \phi }v_{\phi 1}\right] =-B_{0}\left[ \mathbf{\nabla }
\cdot \mathbf{v}_{1}-\frac{\partial }{\partial z}v_{z1}\right]
$$
from $\partial \mathbf{B}_{1}/\partial t=\mathbf{\nabla \times }\left( 
\mathbf{u\times B}_{1}\right) +\mathbf{\nabla \times }\left( \mathbf{v}_{1}
\mathbf{\times B}_{0}\right) $. The linearized momentum equation becomes
\doublespace
\begin{equation}
\begin{array}{c}
\gamma _{0}^{2}W_{0}\left[ \frac{\partial }{\partial t}\mathbf{v}_{1}+
\mathbf{u}\cdot \nabla \mathbf{v}_{1}-\frac{1}{4\pi c^{2}\gamma _{0}^{2}W_{0}
}\left( \frac{\partial }{\partial t}\mathbf{v}_{1}\mathbf{\times B}
_{0}\right) \mathbf{\times B}_{0}\right] = \\ 
-\mathbf{\bigtriangledown }P_{1}-\frac{\mathbf{u}}{c^{2}}\frac{\partial }{
\partial t}P_{1}+\frac{1}{4\pi }\left[ (\bigtriangledown \times \mathbf{B}
_{0})\times \mathbf{B}_{1}+(\bigtriangledown \times \mathbf{B}_{1})\times 
\mathbf{B}_{0}\right] +\frac{1}{4\pi c^{2}}\left( \mathbf{u\times }\frac{
\partial }{\partial t}\mathbf{B}_{1}\right) \mathbf{\times B}_{0}
\label{A9}
\end{array}
\end{equation}
\baselineskip 12pt
where I have used
$$
\frac{\mathbf{j}_{0}\mathbf{\times B}_{1}}{c}=\frac{(\mathbf{\nabla \times B}
_{0})\mathbf{\times B}_{1}}{4\pi }~,
$$
and
$$
\frac{\mathbf{j}_{1}\mathbf{\times B}_{0}}{c}=\frac{(\mathbf{\nabla \times B}
_{1})\mathbf{\times B}_{0}}{4\pi }-\frac{1}{4\pi c}\frac{\partial }{\partial
t}\mathbf{E}_{1}\mathbf{\times B}_{0}=\frac{(\mathbf{\nabla \times B}_{1})
\mathbf{\times B}_{0}}{4\pi }+\frac{1}{4\pi c^{2}}\left( \mathbf{u}\times 
\frac{\partial }{\partial t}\mathbf{B}_{1}+\frac{\partial }{\partial t}
\mathbf{v}_{1}\mathbf{\times B}_{0}\right)\mathbf{\times B}_{0}~,
$$
which includes the displacement current. The components of the linearized
momentum equation can be written as
\begin{equation} \eqnum{A10a}
\gamma _{0}^{2}W_{0}\left[ \left( 1+\frac{V_{A}^{2}}{\gamma _{0}^{2}c^{2}}
\right) \frac{\partial }{\partial t}v_{r1}+u\frac{\partial }{\partial z}
v_{r1}\right] =-\frac{\partial }{\partial r}P_{1}+\frac{B_{0}}{4\pi }\left( 
\frac{\partial }{\partial z}B_{r1}+\frac{u}{c^{2}}\frac{\partial }{\partial t
}B_{r1}-\frac{\partial }{\partial r}B_{z1}\right)~,  \label{A10a}
\end{equation}
\begin{equation} \eqnum{A10b}
\gamma _{0}^{2}W_{0}\left[ \left( 1+\frac{V_{A}^{2}}{\gamma _{0}^{2}c^{2}}
\right) \frac{\partial }{\partial t}v_{\phi 1}+u\frac{\partial }{\partial z}
v_{\phi 1}\right] =-\frac{1}{r}\frac{\partial }{\partial \phi }P_{1}+\frac{
B_{0}}{4\pi }\left( \frac{\partial }{\partial z}B_{\phi 1}+\frac{u}{c^{2}}
\frac{\partial }{\partial t}B_{\phi 1}-\frac{1}{r}\frac{\partial }{\partial
\phi }B_{z1}\right)~,  \label{A10b}
\end{equation}
and
\begin{equation} \eqnum{A10c}
\gamma _{0}^{2}W_{0}\left[ \frac{\partial }{\partial t}v_{z1}+u\frac{
\partial }{\partial z}v_{z1}\right] =-\frac{\partial }{\partial z}P_{1}-
\frac{u}{c^{2}}\frac{\partial }{\partial t}P_{1}  \label{A10c}
\end{equation}
\setcounter{equation}{10}
where $V_{A}^{2}\equiv B_{0}^{2}/(4\pi W_{0})$.

\vspace{-0.5cm}
\section{Normal Mode Dispersion Relation}

In cylindrical geometry perturbations $\rho _{1}$,
$\bf{v}_{1}$, $P_{1}$, and $\bf{B}_{1}$ can be considered to consist
of Fourier components of the form 
$$
f_{1}(r,\phi ,z,t)=f_{1}(r)e^{i(kz\pm n\phi -\omega t)}
$$
where the flow is in the $z$ direction, and $r$ is in the radial direction
with the jet bounded by $r=R$. In cylindrical geometry $n$, an integer, is
the azimuthal wavenumber, for $n>0$ waves propagate at an angle to the flow
direction, where $+n$ and $-n$ refer to wave propagation in the clockwise and
counterclockwise sense, respectively, when viewed outwards along the flow
direction. In general the goal is to write a differential equation for the
radial dependence of the total pressure perturbation $P_{1}^{\ast }\equiv
P_{1}+(\mathbf{B}_{1}\cdot \mathbf{B}_{0})/4\pi =P_{1}^{\ast }(r)\exp
[i(kz\pm n\phi -\omega t)]$.  The differential equation can be obtained from the
energy equation by using the momentum equation and writing the velocity
components $v_{r1},v_{\phi 1},v_{z1}$ in terms of $P_{1}^{\ast },u,B_{0}$.
The components of the linearized momentum equation (eqs.\ A10a, b, \&
c) written in the form
\begin{equation} \eqnum{B1a}
\gamma _{0}^{2}W_{0}\left[ \left( 1+\frac{V_{A}^{2}}{\gamma _{0}^{2}c^{2}}
\right) \frac{\partial }{\partial t}v_{r1}+u\frac{\partial }{\partial z}
v_{r1}\right] =-\frac{\partial }{\partial r}P_{1}^{\ast }+\frac{B_{0}}{4\pi }
\left( \frac{\partial }{\partial z}B_{r1}+\frac{u}{c^{2}}\frac{\partial }{
\partial t}B_{r1}\right)~,  \label{B1a}
\end{equation}
\begin{equation} \eqnum{B1b}
\gamma _{0}^{2}W_{0}\left[ \left( 1+\frac{V_{A}^{2}}{\gamma _{0}^{2}c^{2}}
\right) \frac{\partial }{\partial t}v_{\phi 1}+u\frac{\partial }{\partial z}
v_{\phi 1}\right] =-\frac{1}{r}\frac{\partial }{\partial \phi }P_{1}^{\ast }+
\frac{B_{0}}{4\pi }\left( \frac{\partial }{\partial z}B_{\phi 1}+\frac{u}{
c^{2}}\frac{\partial }{\partial t}B_{\phi 1}\right)~,  \label{B1b}
\end{equation}
and
\begin{equation} \eqnum{B1c}
\gamma _{0}^{2}W_{0}\left[ \frac{\partial }{\partial t}v_{z1}+u\frac{
\partial }{\partial z}v_{z1}\right] =-\left( \frac{\partial }{\partial z}
P_{1}^{\ast }+\frac{u}{c^{2}}\frac{\partial }{\partial t}P_{1}^{\ast
}\right) +\frac{B_{0}}{4\pi }\left( \frac{\partial }{\partial z}B_{z1}+\frac{
u}{c^{2}}\frac{\partial }{\partial t}B_{z1}\right)  \label{B1c}
\end{equation}
\setcounter{equation}{1}
along with
$$
\frac{\partial }{\partial t}\mathbf{B}_{1}=-c(\mathbf{\nabla }\times \mathbf{
E}_{1})=\mathbf{\nabla \times }(\mathbf{u\times B}_{1})+\mathbf{\nabla
\times }(\mathbf{v}_{1}\mathbf{\times B}_{0})
$$
are used to provide relations between $B_{r1}$ and $v_{r1}$, and
$B_{\phi 1}$ and $v_{\phi 1}$
\begin{equation} \eqnum{B2a}
\frac{\partial }{\partial t}B_{r1}+u\frac{\partial }{\partial z}B_{r1}=B_{0}
\frac{\partial }{\partial z}v_{r1}~,  \label{B2a}
\end{equation}
\begin{equation} \eqnum{B2b}
\frac{\partial }{\partial t}B_{\phi 1}+u\frac{\partial }{\partial z}B_{\phi
1}=B_{0}\frac{\partial }{\partial z}v_{\phi 1}~,  \label{B2b}
\end{equation}
and to provide a relation between $B_{z1}$, $v_{z1}$, and $P_{1}^{\ast}$
\doublespace
\begin{equation} \eqnum{B2c}
\begin{array}{c}
\left[ 1+\frac{V_{A}^{2}}{c^{2}}\left( \frac{1}{\tilde{a}^{2}}+\frac{\Gamma 
}{\Gamma -1}\frac{1}{c^{2}}\right) \right] \left[ \frac{\partial }{\partial t
}B_{z1}+u\frac{\partial }{\partial z}B_{z1}\right] -\frac{V_{A}^{2}}{\gamma
_{0}^{2}c^{2}}\frac{\partial }{\partial t}B_{z1}= 
\\
B_{0}\frac{\partial }{\partial z}v_{z1}+2B_{0}\gamma _{0}^{2}\frac{u}{c^{2}}
\left[ \frac{\partial }{\partial t}v_{z1}+u\frac{\partial }{\partial z}v_{z1}
\right]
\\  +\frac{B_{0}}{W_{0}}\left( \frac{1}{\tilde{a}^{2}}+\frac{\Gamma }{
\Gamma -1}\frac{1}{c^{2}}\right) \left[ \frac{\partial }{\partial t}
P_{1}^{\ast }+u\frac{\partial }{\partial z}P_{1}^{\ast }\right] -\frac{B_{0}
}{c^{2}\gamma _{0}^{2}W_{0}}\frac{\partial }{\partial t}P_{1}^{\ast}~.
\label{B2c}
\end{array}
\end{equation}
\baselineskip 12pt
\setcounter{equation}{2}
To obtain equation (B2c) I have used
$$
\frac{\partial }{\partial t}B_{z1}+u\frac{\partial }{\partial z}B_{z1}=-
\frac{B_{0}}{r}\left[ \frac{\partial }{\partial r}(rv_{r1})+\frac{\partial }{
\partial \phi }v_{\phi 1}\right] =-B_{0}\left[ \mathbf{\bigtriangledown }
\cdot \mathbf{v}_{1}-\frac{\partial }{\partial z}v_{z1}\right]
$$
where
$$
-\gamma _{0}^{2}W_{0}\mathbf{\bigtriangledown }\cdot \mathbf{v}_{1}=\gamma
_{0}^{2}\left( \frac{1}{\tilde{a}^{2}}+\frac{\Gamma }{\Gamma -1}\frac{1}{
c^{2}}\right) \left[ \frac{\partial }{\partial t}P_{1}+u\frac{\partial }{
\partial z}P_{1}\right] -\frac{1}{c^{2}}\frac{\partial }{\partial t}
P_{1}+2\gamma _{0}^{4}\frac{u}{c^{2}}W_{0}\left[ \frac{\partial }{\partial t}
v_{z1}+u\frac{\partial }{\partial z}v_{z1}\right]
$$
from the energy equation (eq.\ A8), and
$$
W_{1}=\rho _{1}+\frac{\Gamma }{\Gamma -1}\frac{P_{1}}{c^{2}}=\left( \frac{1}{
\tilde{a}^{2}}+\frac{\Gamma }{\Gamma -1}\frac{1}{c^{2}}\right) P_{1}~.
$$
Using equations (B1a, b, \& c) combined with
\doublespace
$$ 
\begin{array}{c}
\frac{\partial }{\partial t}f_{1}(r,\phi ,z,t) =-i\omega
f_{1}(r)e^{i\left( kz\pm n\phi -\omega t\right) } \\
\frac{\partial }{\partial r}f_{1}(r,\phi ,z,t) =\frac{\partial }{\partial r
}f_{1}(r)e^{i\left( kz\pm n\phi -\omega t\right) } \\
\frac{\partial }{\partial \phi }f_{1}(r,\phi ,z,t) =\pm
inf_{1}(r)e^{i\left( kz\pm n\phi -\omega t\right) } \\
\frac{\partial }{\partial z}f_{1}(r,\phi ,z,t) =+ikf_{1}(r)e^{i\left(
kz\pm n\phi -\omega t\right) }
\end{array}
$$
\baselineskip 12pt
allows the velocity components to be written as
\begin{equation} \eqnum{B3a}
i\gamma _{0}^{2}W_{0}\left[ ku-\omega \left( 1+\frac{V_{A}^{2}}{\gamma
_{0}^{2}c^{2}}\right) \right] v_{r1}=-\frac{\partial }{\partial r}
P_{1}^{\ast }+i\frac{B_{0}}{4\pi }\left( k-\omega \frac{u}{c^{2}}\right)
B_{r1}~,  \label{B3a}
\end{equation}
\begin{equation} \eqnum{B3b}
i\gamma _{0}^{2}W_{0}\left[ ku-\omega \left( 1+\frac{V_{A}^{2}}{\gamma
_{0}^{2}c^{2}}\right) \right] v_{\phi 1}=-\frac{1}{r}\frac{\partial }{
\partial \phi }P_{1}^{\ast }+i\frac{B_{0}}{4\pi }\left( k-\omega \frac{u}{
c^{2}}\right) B_{\phi 1}~,  \label{B3b}
\end{equation}
and
\begin{equation} \eqnum{B3c}
\gamma _{0}^{2}W_{0}\left[ ku-\omega \right] v_{z1}=-\left( k-\omega \frac{u
}{c^{2}}\right) P_{1}^{\ast }+\frac{B_{0}}{4\pi }\left( k-\omega \frac{u}{
c^{2}}\right) B_{z1}~.  \label{B3c}
\end{equation}
\setcounter{equation}{3}
The perturbed magnetic field components from equations (B2a, b,\& c)
become
\begin{equation} \eqnum{B4a}
B_{r1}=\frac{kv_{r1}}{ku-\omega }B_{0}~,  \label{B4a}
\end{equation}
\begin{equation} \eqnum{B4b}
B_{\phi 1}=\frac{kv_{\phi 1}}{ku-\omega }B_{0}~,  \label{B4b}
\end{equation}
and
\begin{equation} \eqnum{B4c}
B_{z1}=\frac{\frac{1}{W_{0}}\left[ \frac{\left( ku-\omega \right) }{a^{2}}
-\left( k-\omega u/c^{2}\right) \frac{u}{c^{2}}\right] P_{1}^{\ast }+\gamma
_{0}^{2}\left[ \left( k-\omega u/c^{2}\right) +\left( ku-\omega \right) 
\frac{u}{c^{2}}\right] v_{z1}}{\left( ku-\omega \right) +V_{A}^{2}\left[ 
\frac{\left( ku-\omega \right) }{a^{2}}-\left( k-\omega u/c^{2}\right) \frac{
u}{c^{2}}\right] }B_{0}  \label{B4c}
\end{equation}
\setcounter{equation}{4}
where I have used 
\doublespace
$$
\begin{array}{c}
 k+2\gamma _{0}^{2}\left( ku-\omega \right)
u/c^{2}=\gamma _{0}^{2}\left[ \left( k-\omega u/c^{2}\right) +\left(
ku-\omega \right) u/c^{2}\right]~, \\ 
\gamma _{0}^{2}\left( ku-\omega
\right) \left[ \tilde{a}^{-2}+\Gamma \left( \Gamma -1\right) ^{-1}c^{-2}
\right] +\omega /c^{2}=\gamma _{0}^{2}\left[ \left( ku-\omega \right)
/a^{2}-\left( k-\omega u/c^{2}\right) u/c^{2}\right]~,
\end{array}
$$
\baselineskip 12pt
and
$$ 
a^{2}\equiv \left( \frac{1}{\tilde{a}^{2}}+\frac{\Gamma }{\Gamma -1}\frac{1}{
c^{2}}-\frac{1}{c^{2}}\right) ^{-1}=\frac{\Gamma P}{\rho +\frac{\Gamma }{
\Gamma -1}\frac{P}{c^{2}}}
$$
to obtain the expression for $B_{z1}$. Using equations (B4a, b, \& c) for the
perturbed magnetic field components, I obtain the following relations
between the perturbed velocity components $\mathbf{v}_{1}$ and the total
pressure perturbation $P_{1}^{\ast }$:
\begin{equation} \eqnum{B5a}
v_{r1}\equiv C_{r}\frac{\partial }{\partial r}P_{1}^{\ast }=i\frac{1}{X}
\frac{\partial }{\partial r}P_{1}^{\ast }=i\frac{\left( ku-\omega \right) }{
\gamma _{0}^{2}W_{0}\gamma _{A}^{2}\left[ \left( ku-\omega \right)
^{2}-\left( k-\omega u/c^{2}\right) ^{2}v_{A}^{2}\right] }\frac{\partial }{
\partial r}P_{1}^{\ast }~,  \label{B5a}
\end{equation}
\begin{equation} \eqnum{B5b}
v_{\phi 1}\equiv C_{\phi }P_{1}^{\ast }=\mp \frac{n}{r}\frac{1}{X}
P_{1}^{\ast }=\mp \frac{n}{r}\frac{\left( ku-\omega \right) }{\gamma
_{0}^{2}W_{0}\gamma _{A}^{2}\left[ \left( ku-\omega \right) ^{2}-\left(
k-\omega u/c^{2}\right) ^{2}v_{A}^{2}\right] }P_{1}^{\ast }~,
\label{B5b}
\end{equation}
and
\begin{equation} \eqnum{B5c}
v_{z1}\equiv C_{z}P_{1}^{\ast }=-\frac{\left( ku-\omega \right) \left(
k-\omega u/c^{2}\right) }{\gamma _{0}^{2}W_{0}\left\{ \left( ku-\omega
\right) ^{2}+\gamma _{A}^{2}v_{A}^{2}\left[ \frac{\left( ku-\omega \right)
^{2}}{a^{2}}-\left( k-\omega u/c^{2}\right) ^{2}\right] \right\} }
P_{1}^{\ast }~.  \label{B5c}
\end{equation}
\setcounter{equation}{5}
To obtain the above relationships I have used
$$
\left( ku-\omega \right) -\frac{V_{A}^{2}}{\gamma _{0}^{2}}\left( \frac{
k-\omega u/c^{2}}{ku-\omega }k+\frac{\omega }{c^{2}}\right) =\gamma _{A}^{2}
\left[ \left( ku-\omega \right) -\frac{\left( k-\omega u/c^{2}\right) ^{2}}{
\left( ku-\omega \right) }v_{A}^{2}\right]
$$
in addition to 
$$\gamma _{0}^{2}\left( ku-\omega \right) \left[ \tilde{a}
^{-2}+\Gamma \left( \Gamma -1\right) ^{-1}c^{-2}\right] +\omega
/c^{2}=\gamma _{0}^{2}\left[ \left( ku-\omega \right) /a^{2}-\left( k-\omega
u/c^{2}\right) u/c^{2}\right] 
$$
where 
$$
v_{A}^{2}\equiv \frac{V_{A}^{2}}{1+V_{A}^{2}/c^{2}}
$$
is the Alfv\'{e}n wave speed and
$
\gamma _{A}^{2}=\left( 1-v_{A}^{2}/c^{2}\right) ^{-1}
$
is an Alfv\'{e}nic Lorentz factor.  Note that $V_{A}^{2}=\gamma
_{A}^{2}v_{A}^{2}$ and $\gamma _{A}^{2}=1+V_{A}^{2}/c^{2}$. Thus we
have that
\doublespace
\begin{equation}
\begin{array}{c}
\mathbf{\nabla }\cdot \mathbf{v}_{1}=C_{r}\frac{\partial ^{2}}{
\partial r^{2}}P_{1}^{\ast }+\frac{C_{r}}{r}\frac{\partial }{\partial r}
P_{1}^{\ast }+\frac{C_{\phi }}{r}\frac{\partial }{\partial \phi }P_{1}^{\ast
}+C_{z}\frac{\partial }{\partial z}P_{1}^{\ast } \\ 
=\frac{i}{X}\frac{\partial ^{2}}{\partial r^{2}}P_{1}^{\ast }+\frac{1}{r}
\frac{i}{X}\frac{\partial }{\partial r}P_{1}^{\ast }-\frac{n^{2}}{r}\frac{i}{
X}P_{1}^{\ast }+ikC_{z}P_{1}^{\ast }~.
\end{array}
\label{B6}
\end{equation}
\baselineskip 12pt
Using the energy equation (eq.\ A8) written in the form
\doublespace
$$
\begin{array}{c}
-\gamma _{0}^{2}W_{0}\mathbf{\bigtriangledown }\cdot \mathbf{v}_{1}=\gamma
_{0}^{2}\left( \frac{1}{\tilde{a}^{2}}+\frac{\Gamma }{\Gamma -1}\frac{1}{
c^{2}}\right) \left[ \frac{\partial }{\partial t}P_{1}^{\ast }+u\frac{
\partial }{\partial z}P_{1}^{\ast }\right] -\frac{1}{c^{2}}\frac{\partial }{
\partial t}P_{1}^{\ast } \\ 
-\gamma _{0}^{2}\left( \frac{1}{\tilde{a}^{2}}+\frac{\Gamma }{\Gamma -1}
\frac{1}{c^{2}}\right) \left[ \frac{\partial }{\partial t}B_{z1}+u\frac{
\partial }{\partial z}B_{z1}\right] +\frac{1}{c^{2}}\frac{\partial }{
\partial t}B_{z1}+2\gamma _{0}^{4}\frac{u}{c^{2}}W_{0}\left[ \frac{\partial 
}{\partial t}v_{z1}+u\frac{\partial }{\partial z}v_{z1}\right]
\end{array}~,
$$
\baselineskip 12pt
inserting
$$
\frac{\partial }{\partial t}B_{z1}+u\frac{\partial }{\partial z}B_{z1}=-B_{0}
\left[ \mathbf{\bigtriangledown }\cdot \mathbf{v}_{1}-\frac{\partial }{
\partial z}v_{z1}\right]~,
$$
and using $v_{z1}=C_{z}P_{1}^{\ast }$ gives
\doublespace
\begin{equation}
\begin{array}{c}
\mathbf{\nabla }\cdot \mathbf{v}_{1}=-i\frac{Y}{\gamma
_{0}^{2}W_{0}}\left[ 1+\frac{V_{A}^{2}}{\gamma _{0}^{2}}\frac{Y}{\left(
ku-\omega \right) }\right] ^{-1}P_{1}^{\ast } \\
-i\left[ 1+\frac{V_{A}^{2}}{
\gamma _{0}^{2}}\frac{Y}{\left( ku-\omega \right) }\right] ^{-1}\left[
2\gamma _{0}^{2}\left( ku-\omega \right) \frac{u}{c^{2}}-\frac{V_{A}^{2}}{
\gamma _{0}^{2}}Y\frac{k}{\left( ku-\omega \right) }\right] C_{z}P_{1}^{\ast
}  \label{B7}
\end{array}
\end{equation}
\baselineskip 12pt
where
\begin{equation}
Y=\gamma _{0}^{2}\left[ \left( ku-\omega \right) /a^{2}-\left( k-\omega
u/c^{2}\right) u/c^{2}\right]~.  \label{B8}
\end{equation}
Setting equations (B6)
and (B7) equal gives us a differential equation for $P_{1}^{\ast }$ in the
form of Bessel's equation
\begin{equation}
r^{2}\frac{\partial ^{2}}{\partial r^{2}}P_{1}^{\ast }+r\frac{\partial }{
\partial r}P_{1}^{\ast }+\left[ \beta ^{2}r^{2}-n^{2}\right] P_{1}^{\ast }=0
\label{B9}
\end{equation}
where
\doublespace
\begin{equation}
\begin{array}{c}
\beta ^{2}=\frac{YX}{\gamma _{0}^{2}W_{0}}\left[ 1+\frac{V_{A}^{2}}{\gamma
_{0}^{2}}\frac{Y}{\left( ku-\omega \right) }\right] ^{-1} \\
+kXC_{z}+\left[ 1+
\frac{V_{A}^{2}}{\gamma _{0}^{2}}\frac{Y}{\left( ku-\omega \right) }\right]
^{-1}\left[ 2\gamma _{0}^{2}\left( ku-\omega \right) \frac{u}{c^{2}}-\frac{
V_{A}^{2}}{\gamma _{0}^{2}}Y\frac{k}{\left( ku-\omega \right) }\right] XC_{z}
~.  \label{B10}
\end{array}
\end{equation}
\baselineskip 12pt

I can simplify the expression for $\beta ^{2}$ by writing 
$$
\beta ^{2}=\frac{X}{\left( ku-\omega \right) +\frac{V_{A}^{2}}{\gamma
_{0}^{2}}Y}\left\{ \frac{\left( ku-\omega \right) }{\gamma _{0}^{2}W_{0}}Y+
\left[ \left( ku-\omega \right) +\frac{V_{A}^{2}}{\gamma _{0}^{2}}Y\right]
kC_{z}+\left[ 2\gamma _{0}^{2}\left( ku-\omega \right) ^{2}\frac{u}{c^{2}}
\right] C_{z}-\frac{V_{A}^{2}}{\gamma _{0}^{2}}kYC_{z}\right\}
$$
from which it follows that
$$
\beta ^{2}=\left\{ \frac{X}{\left( ku-\omega \right) +\frac{V_{A}^{2}}{
\gamma _{0}^{2}}Y}\right\} \times \left\{ \frac{\left( ku-\omega \right) }{
\gamma _{0}^{2}W_{0}}Y+\gamma _{0}^{2}\left( ku-\omega \right) \left[ \left(
k-\omega u/c^{2}\right) +\left( ku-\omega \right) u/c^{2}\right]
C_{z}\right\}
$$
where I have used $2\gamma _{0}^{2}\left( ku-\omega \right) u/c^{2}=\gamma
_{0}^{2}\left[ \left( k-\omega u/c^{2}\right) +\left( ku-\omega \right)
u/c^{2}\right] -k$.\ Substituting the expressions for X and C$_{z}$ from
equations (B5a, b, \& c), and Y from equation (B8) and modest algebraic manipulation
yields
\doublespace
\begin{equation}
\begin{array}{c}
\beta ^{2}=\left\{ \frac{\gamma _{0}^{2}\gamma _{A}^{2}\left[ \left(
ku-\omega \right) ^{2}-\left( k-\omega u/c^{2}\right) ^{2}v_{A}^{2}\right] }{
\left( a^{2}+\gamma _{A}^{2}v_{A}^{2}\right) \left( ku-\omega \right)
^{2}+\gamma _{A}^{2}v_{A}^{2}a^{2}\left( ku-\omega \right) \left( k-\omega
u/c^{2}\right) u/c^{2}}\right\} \times \\ 
\left\{ \left( ku-\omega \right) ^{2}+\left( ku-\omega \right) \left(
k-\omega u/c^{2}\right) a^{2}u/c^{2}-\frac{\left( ku-\omega \right) ^{2}
\left[ \left( k-\omega u/c^{2}\right) ^{2}a^{2}+\left( ku-\omega \right)
\left( k-\omega u/c^{2}\right) ua^{2}/c^{2}\right] }{\left( a^{2}+\gamma
_{A}^{2}v_{A}^{2}\right) \left( ku-\omega \right) ^{2}-\gamma
_{A}^{2}v_{A}^{2}a^{2}\left( k-\omega u/c^{2}\right) ^{2}}\right\}~.
\end{array} \label{B11}
\end{equation}
\baselineskip 12pt
Additional regrouping provides the following form
\doublespace
\begin{equation}
\begin{array}{c}
\beta ^{2}=\left\{ \frac{\gamma _{0}^{2}\gamma _{A}^{2}\left[ \left(
ku-\omega \right) ^{2}-\left( k-\omega u/c^{2}\right) ^{2}v_{A}^{2}\right] }{
\left( a^{2}+\gamma _{A}^{2}v_{A}^{2}\right) \left( ku-\omega \right)
^{2}+\gamma _{A}^{2}v_{A}^{2}a^{2}\left( ku-\omega \right) \left( k-\omega
u/c^{2}\right) u/c^{2}}\right\} \times \\ 
\left\{ \frac{\left[ \left( a^{2}+\gamma _{A}^{2}v_{A}^{2}\right) \left(
ku-\omega \right) ^{2}+\gamma _{A}^{2}v_{A}^{2}a^{2}\left( ku-\omega \right)
\left( k-\omega u/c^{2}\right) u/c^{2}\right] \left[ \left( ku-\omega
\right) ^{2}-\left( k-\omega u/c^{2}\right) ^{2}a^{2}\right] }{\left(
a^{2}+\gamma _{A}^{2}v_{A}^{2}\right) \left( ku-\omega \right) ^{2}-\gamma
_{A}^{2}v_{A}^{2}a^{2}\left( k-\omega u/c^{2}\right) ^{2}}\right\}
\end{array} \label{B12}
\end{equation}
\baselineskip 12pt
from which I find that $\beta ^{2}$ can be written in the
compact form:
\begin{equation}
\beta ^{2}\equiv \left[ \frac{\gamma _{0}^{2}\left( \varpi
^{2}-\kappa ^{2}a^{2}\right) \left( \varpi ^{2}-\kappa ^{2}v_{A}^{2}\right) 
}{v_{ms}^{2} \varpi ^{2}-\kappa ^{2}v_{A}^{2}a^{2}}\right]~.  \label{B13}
\end{equation}
where $\varpi ^{2}\equiv \left( \omega -ku\right) ^{2}$, $\kappa
^{2}\equiv \left( k-\omega u/c^{2}\right) ^{2}$, and where the fast
magnetosonic speed perpendicular to the magentic field is given by
(e.g., Vlahakis \& K\"onigl 2003)
$$
v_{ms} \equiv \left[a^2 + v_A^2 - a^2v_A^2/c^2\right]^{1/2} = \left[a^2/\gamma_A^2 + v_A^2\right]^{1/2}~.
$$ 
It is easily seen that this
expression for $\beta ^{2}$ reduces to the relativistic pure fluid form
$$
\beta ^{2}\longrightarrow \left[ \frac{\gamma _{0}^{2}\left( \varpi
^{2}-\kappa ^{2}a^{2}\right) }{a^{2}}\right] =\gamma _{0}^{2}\left[ \frac{
\left( ku-\omega \right) ^{2}}{a^{2}}-\left( k-\omega u/c^{2}\right) ^{2}
\right]
$$
given in Hardee (2000) and that this expression for $\beta ^{2}$ reduces to
the non-relativistic MHD form
$$
\beta ^{2}\longrightarrow \left[ \frac{\left( \varpi ^{2}-\kappa
^{2}a^{2}\right) \left( \varpi ^{2}-\kappa ^{2}v_{A}^{2}\right) }{\left(
a^{2}+v_{A}^{2}\right) \varpi ^{2}-\kappa ^{2}v_{A}^{2}a^{2}}\right] =\left[ 
\frac{\left( ku-\omega \right) ^{4}}{\left( a^{2}+V_{A}^{2}\right) \left(
ku-\omega \right) ^{2}-k^{2}V_{A}^{2}a^{2}}-k^{2}\right]
$$
where $\kappa ^{2}\longrightarrow k^{2}$\ and $v_{A}^{2}\longrightarrow
V_{A}^{2}$ given in Hardee, Clarke \& Rosen (1997).

The solutions that are well behaved at jet center and at infinity are $
P_{j1}^{\ast }(r\leq R)=C_{j}J_{\pm n}(\beta _{j}r)$, and $P_{e1}^{\ast
}(r\geq R)=C_{e}H_{\pm n}^{(1)}(\beta _{e}r)$, respectively, where
$J_{\pm n}$ and $H_{\pm n}^{(1)}$ are the Bessel and Hankel
functions with arguments defined as
\begin{equation} \eqnum{B14a}
\beta _{j}^{2}\equiv \left[ \frac{\gamma _{j}^{2}\left(
\varpi _{j}^{2}-\kappa _{j}^{2}a_{j}^{2}\right) \left( \varpi
_{j}^{2}-\kappa _{j}^{2}v_{Aj}^{2}\right) }{v_{msj}^{2} \varpi _{j}^{2}-\kappa
_{j}^{2}v_{Aj}^{2}a_{j}^{2}}\right]~,  \label{B14a}
\end{equation}
and
\begin{equation} \eqnum{B14b}
\beta _{e}^{2}\equiv \left[ \frac{\gamma _{e}^{2}\left(
\varpi _{e}^{2}-\kappa _{e}^{2}a_{e}^{2}\right) \left( \varpi
_{e}^{2}-\kappa _{e}^{2}v_{Ae}^{2}\right) }{v_{mse}^{2} \varpi _{e}^{2}-\kappa
_{e}^{2}v_{Ae}^{2}a_{e}^{2}}\right]~,  \label{B14b}
\end{equation}
\setcounter{equation}{14}
where $\varpi _{j,e}^{2}\equiv \left( \omega -ku_{j,e}\right) ^{2}$, $\kappa
_{j,e}^{2}\equiv \left( k-\omega u_{j,e}/c^{2}\right) ^{2}$, $\gamma
_{j,e}^{2}\equiv \left( 1-u_{j,e}^{2}/c^{2}\right) ^{-1}$ and $\gamma
_{Aj,e}^{2}\equiv \left( 1-v_{Aj,e}^{2}/c^{2}\right) ^{-1}$. The jet flow
speed and external flow speed are positive if flow is in the $+z$ direction.

The condition that the total pressure be continuous across the jet boundary
requires that
\begin{equation}
C_{j}J_{\pm n}(\beta _{j}R)=C_{e}H_{\pm n}^{(1)}(\beta _{e}R)~. \label{B15}
\end{equation}
The first derivative of the total pressure is given by
$$
\frac{\partial }{\partial r}P_{1}^{\ast }=-iXv_{r1}~.
$$
and with 
$$
v_{r1}\equiv \left[ \frac{\partial }{\partial t}+u\cdot \nabla \right] \xi
_{r}=-i\left( \omega -ku\right) \xi _{r}
$$
where $\xi _{r}$\ is the fluid displacement in the radial direction it
follows that
\begin{equation}
\frac{\partial P_{1}^{\ast }}{\partial r}=-\left( \omega -ku\right) X\xi _{r}
~. \label{B16}
\end{equation}
The radial displacement of the jet and external medium must be equal at the
jet boundary, i.e., $\xi _{r}^{j}(R)=\xi _{r}^{e}(R)$, from which it follows
that
\begin{equation}
\frac{\beta _{j}}{-\left( \omega -ku_{j}\right) X_{j}}C_{j}\left. \frac{
\partial J_{n}(\beta _{j}r)}{\partial \left( \beta _{j}r\right) }\right|
_{r=R}=\frac{\beta _{e}}{-\left( \omega -ku_{e}\right) X_{e}}C_{e}\left. 
\frac{\partial H_{n}^{(1)}(\beta _{e}r)}{\partial \left( \beta _{e}r\right) }
\right| _{r=R}~. \label{B17}
\end{equation}
Inserting $C_{j}$ and $C_{e}$ in terms of the Bessel and Hankel functions
leads to a dispersion relation describing the propagation of Fourier
components which can be written in the following form:
\begin{equation}
\frac{\beta _{j}}{\chi _{j}}\frac{J_{n}^{^{\prime }}(\beta _{j}R)}{
J_{n}(\beta _{j}R)}=\frac{\beta _{e}}{\chi _{e}}\frac{H_{n}^{(1)^{\prime
}}(\beta _{e}R)}{H_{n}^{(1)}(\beta _{e}R)}~.  \label{B18}
\end{equation}
where the primes denote derivatives of the Bessel and Hankel functions with
respect to their arguments.  The expressions
\begin{equation} \eqnum{B19a}
\chi _{j}\equiv \gamma _{j}^{2}\gamma _{Aj}^{2}W_{j}\left( \varpi
_{j}^{2}-\kappa _{j}^{2}v_{Aj}^{2}\right)  \label{B19a}
\end{equation}
and
\begin{equation} \eqnum{B19b}
\chi _{e}\equiv \gamma _{e}^{2}\gamma _{Ae}^{2}W_{e}\left( \varpi
_{e}^{2}-\kappa _{e}^{2}v_{Ae}^{2}\right)  \label{B19b}
\end{equation}
readily reduce to the non-relativistic form $\chi =\rho _{0}[(\omega
-ku)^{2}-k^{2}V_{A}^{2}]$ where $W_{0}\longrightarrow \rho _{0}$ given in
Hardee, Clarke \& Rosen (1997). This dispersion relation describes the
normal modes of a cylindrical jet where n = 0, 1, 2, 3, 4, etc. involve
pinching, helical, elliptical, triangular, rectangular, etc. normal mode
distortions of the jet, respectively.

\vspace{-0.5cm}
\section{Analytic Solutions and Approximations}

Each normal mode $n$ contains a \textit{fundamental/surface} wave and
multiple
\textit{body} wave solutions to the dispersion relation. The
low-frequency limiting form for the fundamental/surface modes are
obtained in the limit where $\omega \longrightarrow 0$ and
$k\longrightarrow 0$ but with $
\omega /k\neq 0$. \ In this limit the dispersion relation for the
fundamental ($n = 0$) and surface ($n > 0$) modes is given by
\begin{equation}
\begin{array}{cc}
\chi _{j}\approx -\frac{1}{2}\chi _{e}\left( \beta _{j}R\right) ^{2}\left[
\ln (\frac{\beta _{e}R}{2})+\frac{\pi }{2}\epsilon -i\frac{\pi }{2}\right] &
n=0
\end{array}
\label{C1}
\end{equation}
and
\begin{equation}
\begin{array}{cc}
\chi _{j}\approx -\chi _{e} & n>0
\end{array}
\label{C2}
\end{equation}
where in this limit $\beta _{e}R\longrightarrow 0$ and $\beta
_{j}R\longrightarrow 0$, and I have used the small argument forms for the
Bessel and Hankel functions to write
$$
\frac{J_{n}^{^{\prime }}(\beta _{j}R)}{J_{n}(\beta _{j}R)}\frac{
H_{n}^{(1)}(\beta _{e}R)}{H_{n}^{(1)^{\prime }}(\beta _{e}R)}\approx \left\{ 
\begin{array}{cc}
-\frac{1}{2}\left( \beta _{e}R\right) \left( \beta _{j}R\right) \left[ \ln (
\frac{\beta _{e}R}{2})+\frac{\pi }{2}\epsilon -i\frac{\pi }{2}\right] & n=0
\\ 
-\beta _{e}/\beta _{j} & n>0
\end{array}
\right.
$$
where $\epsilon $ is Euler's constant.

\vspace{-0.5cm}
\subsection{Fundamental Pinch Mode ($n = 0$ ; $m = 0$) in the low
  frequency limit}

\baselineskip 12pt
In the low frequency limit, dispersion relation solutions for the
fundamental axisymmetric pinch mode are obtained from equation (C1)
\doublespace
\begin{equation}
\begin{array}{c}
\gamma_j^2 \gamma_{Aj}^2W_{j}\left( \varpi _{j}^{2}-\kappa _{j}^{2}v_{Aj}^{2}\right) \approx \\ -\frac{1
}{2}\gamma _{e}^{2}\gamma _{Ae}^{2}W_{e}\left( \varpi _{e}^{2}-\kappa
_{e}^{2}v_{Ae}^{2}\right) \left[ \frac{\gamma_j^2\left( \varpi _{j}^{2}-\kappa
_{j}^{2}a_{j}^{2}\right) \left( \varpi _{j}^{2}-\kappa
_{j}^{2}v_{Aj}^{2}\right) }{v_{msj}^{2} \varpi _{j}^{2}-\kappa
_{j}^{2}v_{Aj}^{2}a_{j}^{2}}\right] R^{2}\left[ \ln (\frac{\beta _{e}R}{2})+
\frac{\pi }{2}\epsilon -i\frac{\pi }{2}\right]~.
\end{array}  \label{C3}
\end{equation}
\baselineskip 12pt
Here we have the trivial solution $\varpi _{j}^{2}-\kappa
_{j}^{2}v_{Aj}^{2}=0$ with $v_{w}^{2}=v_{Aj}^{2}$ and the more interesting
zeroth order solution
\begin{equation}
\varpi _{j}^{2} \approx \kappa _{j}^{2}\left[ \frac{
v_{Aj}^{2}a_{j}^{2}}{v_{msj}^{2}}\right] \label{C4}
\end{equation}
with wave speed in the proper frame given by
\begin{equation}
v_{w}^{2}=\varpi_{j}^{2}/\kappa_{j}^{2} \approx \left[ \frac{v_{Aj}^{2}a_{j}^{2}}{v_{msj}^{2}}\right]~.\label{C5}
\end{equation}
To first order this magnetosonic wave solution (eq.\ C3) can be written as
\begin{equation}
\varpi _{j}^{2}\left[ 1-\delta \right] \simeq \kappa _{j}^{2}a_{j}^{2}\left[
\frac{v_{Aj}^{2}}{v_{msj}^{2}}
-\delta \right]~,  \label{C6}
\end{equation}
where
\begin{equation}
\delta \equiv -\frac{1}{2}\gamma _{e}^{2}\frac{\gamma _{Ae}^{2}}{\gamma_{Aj}^2}\frac{W_{e}}{W_{j}}
\frac{\left( \varpi _{e}^{2}-\kappa _{e}^{2}v_{Ae}^{2}\right) }{v_{msj}^{2} }R^{2}\left[ \ln (\frac{\beta
_{e}R}{2})+\frac{\pi }{2}\epsilon -i\frac{\pi }{2}\right]  \label{C7}
\end{equation}
and $\delta$ is complex. Thus, in the low frequency limit this
fundamental pinch mode ($n=0$) solution consists of a
growing and damped wave pair with wave speed in the observer frame
\begin{equation}
\frac{\omega }{k}=\frac{u_{j}\pm v_{w}}{1\pm v_{w}u_{j}/c^{2}}  \label{C8}
\end{equation}
where
\begin{equation}
v_{w}^{2}\simeq a_{j}^{2}\left\{ \frac{v_{Aj}^{2}}{v_{msj}^{2} }+\delta \left[ \frac{v_{Aj}^{2}}{v_{msj}^{2} }-1
\right] \right\}~.  \label{C9}
\end{equation}
Previous work has shown
the unstable growing solution associated with the backwards moving (in the
jet fluid reference frame) wave.

\vspace{-0.5cm}
\subsection{Surface Modes ($n > 0$ ; $m = 0$) in the low frequency limit}

In the low frequency limit the fundamental dispersion relation
solution for all higher order modes ($n>0$) is most easily obtained
from equation (C2) written in the form
\begin{equation}
\gamma _{j}^{2}W_{j}\left[ (\omega
-ku_{j})^{2}-\frac{V_{Aj}^{2}}{\gamma _{j}^{2}}\left( k^{2}-\omega
^{2}/c^{2}\right) \right] =-\gamma _{e}^{2}W_{e} \left[ (\omega
-ku_{e})^{2}-\frac{V_{Ae}^{2}}{\gamma _{e}^{2}}\left( k^{2}-\omega
^{2}/c^{2}\right) \right] 
\label{C10}
\end{equation}
where I have used $\chi \equiv \gamma _{0}^{2}\gamma _{A}^{2}W_{0}\left(
\varpi ^{2}-\kappa ^{2}v_{A}^{2}\right) =\gamma _{0}^{2}W_{0}\left[ (\omega
-ku)^{2}-\left( k^{2}-\omega ^{2}/c^{2}\right) V_{A}^{2}/\gamma _{0}^{2}
\right] $.  The solution can be put in the form
\begin{equation}
\frac{\omega }{k}=\frac{\left[ \eta u_{j}+u_{e}\right] \pm i\eta ^{1/2}\left[
\left( u_{j}-u_{e}\right) ^{2}-V_{As}^{2}/\gamma _{j}^{2}\gamma _{e}^{2}
\right] ^{1/2}}{(1+V_{Ae}^{2}/\gamma _{e}^{2}c^{2})+\eta
(1+V_{Aj}^{2}/\gamma _{j}^{2}c^{2})}  \label{C11}
\end{equation}
where 
$$
\eta \equiv \frac{\gamma _{j}^{2}W_{j}}{\gamma _{e}^{2}W_{e}}~,
$$
and a ``surface'' Alfv\'{e}n speed is defined by
$$
V_{As}^{2}\equiv \left( \gamma _{Aj}^{2}W_{j}+\gamma _{Ae}^{2}W_{e}\right) 
\frac{B_{j}^{2}+B_{e}^{2}}{4\pi W_{j}W_{e}}~.
$$
The jet is stable to higher order $n>0$ fundamental mode perturbations when
\begin{equation}
\gamma _{j}^{2}\gamma _{e}^{2}\left( u_{j}-u_{e}\right) ^{2} < \gamma
_{Aj}^{2}\gamma _{Ae}^{2}\left( W_{j}/\gamma _{Ae}^{2}+W_{e}/\gamma
_{Aj}^{2}\right) \frac{B_{j}^{2}+B_{e}^{2}}{4\pi W_{j}W_{e}}~.
\label{C12}
\end{equation}

Equation (C11) reduces to
the relativistic fluid expression 
\begin{equation} \eqnum{C13a}
\frac{\omega }{k}=\frac{\eta u_{j}+u_{e}}{1+\eta }\pm \frac{i\eta ^{1/2}}{
1+\eta }\left( u_{j}-u_{e}\right)
\label{C13a}
\end{equation}
given in Hardee \& Hughes (2003) equation (6a) where for pressure balance
and equal adiabatic index in jet and external medium $\eta \longrightarrow
\gamma _{j}^{2}a_{e}/\gamma _{e}^{2}a_{j}$. \ Similarly equation (C11)
reduces to the non-relativistic MHD form
\begin{equation} \eqnum{C13b}
\frac{\omega }{k}=\frac{\eta u_{j}+u_{x}}{1+\eta }\pm \frac{i\eta ^{1/2}}{
1+\eta }\left[ \left( u_{j}-u_{e}\right) ^{2}-V_{As}^{2}\right] ^{1/2}
\label{C13b}
\end{equation}
\setcounter{equation}{13}
given by Hardee \& Rosen (2002) eq.\ (4) where
$V_{As}^{2}\longrightarrow \left( \rho _{j}+\rho _{e}\right) \left(
B_{j}^{2}+B_{e}^{2}\right) /\left( 4\pi \rho _{j}\rho _{e}\right)$ and
$\eta \longrightarrow
\rho _{j}/\rho _{e}$.

\vspace{-0.5cm}
\subsection{Body Modes ($n \geq 0$ ; $m \geq 1$) in the low frequency limit}

In the low frequency limit the real part of the \textit{body} wave solutions
can be obtained in the limit $\omega =0$, $k\neq 0$ where the dispersion
relation can be written in the form
\begin{equation}
\cos \left[ \beta _{j}R-(2n+1)\pi /4\right] \approx \epsilon _{n}\equiv 
\frac{\chi _{e}}{\chi _{j}}\frac{\beta _{e}}{\beta _{j}}\left( \frac{\pi
\beta _{j}R}{2}\right) ^{1/2}J_{n}^{^{\prime }}(\beta _{j}R)\frac{
H_{n}^{(1)}(\beta _{e}R)}{H_{n}^{(1)^{\prime }}(\beta _{e}R)}~.  \label{C14}
\end{equation}
Here I have assumed that the large argument form $J_{n}(\beta
_{j}R)\approx \left( 2/\pi \beta _{j}R\right) ^{1/2}\cos \left[ \beta
_{j}R-(2n+1)\pi /4\right] $ applies. In the absence of a magnetic
field and a flow
surrounding the jet, $\chi _{e}=0$, $\epsilon _{n}=0$, and solutions are
found from $\beta _{j}R-(2n+1)\pi /4=\pm m\pi \pm \pi /2$, where $m$ is an
integer. Provided $\epsilon _{n}<<\pi /2$ and $\theta \approx \cos
^{-1}\epsilon _{n}\approx \pm \left( \pi /2-\epsilon _{n}\right)$, solutions
can be found from $\beta _{j}R-(2n+1)\pi /4=\pm \left[ m\pi +\left( \pi
/2\pm \epsilon _{n}\right) \right]$, where for $\pm \epsilon _{n}$ the plus or
minus sign is for m odd or even, respectively. In the limit $\omega =0$
\begin{equation}
\beta _{j}R\approx \left[ \frac{\gamma _{j}^{2}(u_{j}^{2}-a_{j}^{2})(u_{j}^{2}-v_{Aj}^{2})}{
v_{msj}^2u_{j}^{2}-v_{Aj}^2a_{j}^{2}}\right] ^{1/2}kR~, \label{C15}
\end{equation}
and the solutions are given by
\begin{equation}
kR\approx k_{nm}^{\min }R\equiv \left[ \frac{
v_{msj}^2u_{j}^{2}-v_{Aj}^2a_{j}^{2}}{\gamma
_{j}^{2}(u_{j}^{2}-a_{j}^{2})(u_{j}^{2}-v_{Aj}^{2})}\right]
^{1/2}\times \left[ (n+2m-1/2)\pi /2+(-1)^{m}\epsilon _{n}\right]
\label{C16}
\end{equation}
where I have set $m\longrightarrow m+1$ to be consistent with previous
notation so $m=1$ corresponds to the first body mode. 

In the limit $a_{j}^{2}>>v_{Aj}^{2}$ equation (C16) reduces to the
relativistic purely fluid form found in Hardee \& Hughes (2003)

\begin{equation} \eqnum{C17a}
kR\approx k_{nm}^{\min }R\equiv \frac{\left[ (n+2m-1/2)\pi
/2+(-1)^{m}\epsilon _{n}\right] }{\gamma _{j}\left[ M_{j}^{2}-1\right]
^{1/2}}
\label{C17a}
\end{equation}
where $M_{j}^{2}=u_{j}^{2}/a_{j}^{2}$.  In the limit
$v_{Aj}^{2}>>a_{j}^{2}$ equation (C16) becomes
\begin{equation} \eqnum{C17b}
kR\approx k_{nm}^{\min }R\equiv \frac{\left[ (n+2m-1/2)\pi
/2+(-1)^{m}\epsilon _{n}\right] }{\gamma _{j}\left[ M_{Aj}^{2}-1\right]
^{1/2}}
\label{C17b}
\end{equation}
where $M_{Aj}^{2}=u_{j}^{2}/v_{Aj}^{2}$. 

Equation (C16) reduces to the
non-relativistic MHD form found in Hardee \& Rosen (2002)
\begin{equation} \eqnum{C17c}
kR\approx k_{nm}^{\min }R\equiv \left[ \frac{M_{ms}^{2}}{
1-M_{ms}^{2}/M_{Aj}^{2}M_{j}^{2}}-1\right] ^{-1/2}\times \left[
(n+2m-1/2)\pi /2+(-1)^{m}\epsilon _{n}\right]
\label{C17c}
\end{equation}
\setcounter{equation}{17}
where $M_{ms}^{2}=u_{j}^{2}/(a_{j}^{2}+v_{Aj}^{2})$ and I have used 
$$
 \left[ \frac{\gamma _{j}^{2}(u_{j}^{2}-a_{j}^{2})(u_{j}^{2}-v_{Aj}^{2})}{
v_{msj}^2u_{j}^{2}-v_{Aj}^2a_{j}^{2}}\right] =\gamma
_{j}^{2}\left[ \frac{M_{ms}^{2}}{\frac{v_{msj}^{2}}{a^{2}+v_{Aj}^{2}}-\frac{M_{ms}^{2}}{
M_{Aj}^{2}M_{j}^{2}}}-\frac{1-M_{ms}^{2}/M_{Aj}^{2}M_{j}^{2}}{\frac{v_{msj}^{2}}{a^{2}+v_{Aj}^{2}}-\frac{
M_{ms}^{2}}{M_{Aj}^{2}M_{j}^{2}}}\right]
$$
We note here that there is an error in equations (5) in previous articles in
the treatment of the sign on $\epsilon_{n}$ for even values of $m$.

\vspace{-0.5cm}
\subsection{The Resonance Condition}

The resonance conditions are found by evaluating the transmittance, $
\mathcal{T}$, and reflectance, $\mathcal{R}$, of waves at the jet boundary
where $\mathcal{T} = 1 + \mathcal{R}$. With the dispersion relation written
as 
\begin{equation}
\frac{1}{Z_{j}}\frac{J_{n}^{^{\prime }}(\beta _{j}R)}{J_{n}(\beta _{j}R)}=
\frac{1}{Z_{e}}\frac{H_{n}^{(1)^{\prime }}(\beta _{e}R)}{H_{n}^{(1)}(\beta
_{e}R)}  \label{C18}
\end{equation}
where $Z=\chi /\beta$ with
\begin{equation}
Z=\gamma ^{2}\gamma _{A}^{2}W\left( \varpi ^{2}-\kappa ^{2}v_{A}^{2}\right) 
\left[ \frac{\left( a^{2}+\gamma _{A}^{2}v_{A}^{2}\right) \varpi ^{2}-\gamma
_{A}^{2}\kappa ^{2}v_{A}^{2}a^{2}}{\gamma ^{2}\gamma _{A}^{2}\left( \varpi
^{2}-\kappa ^{2}a^{2}\right) \left( \varpi ^{2}-\kappa ^{2}v_{A}^{2}\right) }
\right] ^{1/2}  \label{C19}
\end{equation}
the reflectance 
\begin{equation}
\mathcal{R}=(Z_{e}-Z_{j})/(Z_{e}+Z_{j})~. \label{C20}
\end{equation}
For a fluid
containing no magnetic field $Z$ is a quantity related to the acoustic
normal impedance (Gill 1965).  When $Z_{e}+Z_{j}\approx 0$, $\mathcal{R}$
and $\mathcal{T}$ are large, and the reflected and transmitted waves have an
amplitude much larger than the incident wave.

\vspace{-0.5cm}
\subsubsection{The Fluid Limit (Alfv\'{e}n speed $\ll$
sound speed)}

For the case of a pure fluid
\begin{equation} \eqnum{C21a}
Z_{e}=\frac{W_{e}\left[ \zeta _{e}^{2}+\gamma _{e}^{2}\kappa
_{e}^{2}/k^{2}\right] a_{e}^{2}}{\zeta _{e}}~,  \label{C21a}
\end{equation}
and
\begin{equation} \eqnum{C21b}
Z_{j}=\frac{W_{j}\left[ \zeta _{j}^{2}+\gamma _{j}^{2}\kappa _{j}^{2}/k^{2}
\right] a_{j}^{2}}{\zeta _{j}}~,  \label{C21b}
\end{equation}
\setcounter{equation}{21}
where $\chi /k^{2}=W\left[ \zeta ^{2}+\gamma ^{2}\kappa
^{2}/k^{2}\right] a^{2}$ and $\zeta \equiv \beta /k$.  For non-relativistic
flows where $(u^{2}/c^{2})(\omega /ku)<<1$, $\gamma \approx 1$, and with
adiabatic indicies $\Gamma _{j}=\Gamma _{e}$ the reflectance
\begin{equation}
\mathcal{R}=\frac{(\zeta _{e}-\zeta _{j})(\zeta _{e}\zeta _{j}-1)}{(\zeta
_{e}+\zeta _{j})(\zeta _{e}\zeta _{j}+1)}  \label{C22}
\end{equation}
and a supersonic resonance (Miles 1957) occurs when $\beta _{e}+\beta
_{j}=k(\zeta _{e}+\zeta _{j})=0$.  This supersonic resonance corresponds to
the maximum growth rate of the normal mode solutions to the dispersion
relation. 

I now generalize the results in Hardee (2000) to include flow
in the external medium relative to the source/observer frame.  Here $
Z_{e}+Z_{j}=0$ becomes
\begin{equation}
\Gamma _{e}\zeta _{j}\chi _{e}+\Gamma _{j}\zeta _{e}\chi _{j}=\Gamma
_{e}\zeta _{j}\left( \gamma _{e}^{2}\varpi _{e}^{2}/k^{2}a_{e}^{2}\right)
+\Gamma _{j}\zeta _{e}\left( \gamma _{j}^{2}\varpi
_{j}^{2}/k^{2}a_{j}^{2}\right) =0~. \label{C23}
\end{equation}
A necessary condition for resonance is $\zeta _{j}<0$ and $\zeta _{e}>0$,
and on the real axis
\begin{equation}
\frac{u_{j}-a_{j}}{1-u_{j}a_{j}/c^{2}}>\frac{\omega }{k}>\frac{u_{e}+a_{e}}{%
1+u_{e}a_{e}/c^{2}}~.  \label{C24}
\end{equation}
It follows that the resonance only exists when
\begin{equation} \eqnum{C25a}
\frac{u_{j}-a_{j}}{1-u_{j}a_{j}/c^{2}}>\frac{u_{e}+a_{e}}{1+u_{e}a_{e}/c^{2}}
\label{C25a}
\end{equation}
or equivalently
\begin{equation} \eqnum{C25b}
\frac{u_{j}-u_{e}}{1-u_{j}u_{e}/c^{2}}>\frac{a_{j}+a_{e}}{1+a_{j}a_{e}/c^{2}}~.
 \label{C25b}
\end{equation}
\setcounter{equation}{25}

To find the resonant solution for the real part of the phase velocity I
solve $\zeta _{j}^{2}=\varepsilon ^{2}\zeta _{e}^{2}$ where here I set $
\varepsilon \equiv (\Gamma _{j}\gamma _{j}^{2}\varpi
_{j}^{2}/k^{2}a_{j}^{2})/\left( \Gamma _{e}\gamma _{e}^{2}\varpi
_{e}^{2}/k^{2}a_{e}^{2}\right) =1$ so that
\begin{equation}
\zeta _{j}^{2}=\gamma _{j}^{2}\left[ \frac{\left( \omega /k-u_{j}\right) ^{2}
}{a_{j}^{2}}-\left( 1-\frac{\omega }{k}\frac{u_{j}}{c^{2}}\right) ^{2}\right]
=\zeta _{e}^{2}=\gamma _{e}^{2}\left[ \frac{\left( \omega /k-u_{e}\right)
^{2}}{a_{e}^{2}}-\left( 1-\frac{\omega }{k}\frac{u_{e}}{c^{2}}\right) ^{2}
\right]~.  \label{C26}
\end{equation}
The resulting quadratic equation can be written in the form
\doublespace
\begin{equation}
\begin{array}{c}
\left[ a_{e}^{2}/(\gamma _{e}^{2}\gamma _{sj}^{2})-a_{j}^{2}/(\gamma
_{j}^{2}\gamma _{se}^{2})\right] \left( \frac{\omega }{k}\right) ^{2}-2\left[
\gamma _{j}^{2}a_{e}^{2}u_{j}/\gamma _{sj}^{2}-\gamma
_{e}^{2}a_{j}^{2}u_{e}/\gamma _{se}^{2}\right] \left( \frac{\omega }{k}
\right) \\ +\left[ \gamma _{j}^{2}a_{e}^{2}\left( u_{j}^{2}-a_{j}^{2}\right)
-\gamma _{e}^{2}a_{j}^{2}\left( u_{e}^{2}-a_{e}^{2}\right) \right] =0~,
\end{array} \label{C27}
\end{equation}
\baselineskip 12pt
where I have used
$$
\left[ \gamma _{j}^{2}a_{e}^{2}\left( 1-\frac{a_{j}^{2}u_{j}^{2}}{c^{4}}
\right) -\gamma _{e}^{2}a_{j}^{2}\left( 1-\frac{a_{e}^{2}u_{e}^{2}}{c^{4}}
\right) \right] =\left[ a_{e}^{2}/(\gamma _{e}^{2}\gamma
_{sj}^{2})-a_{j}^{2}/(\gamma _{j}^{2}\gamma _{se}^{2})\right]~,
$$
and $\gamma _{s}^{2}\equiv \left( 1-a^{2}/c^{2}\right) ^{-1}$.  The
solutions to equation (C27) are given by
\begin{equation}
\frac{\omega }{k}=\frac{\left[ \gamma _{j}^{2}a_{e}^{2}u_{j}/\gamma
_{sj}^{2}-\gamma _{e}^{2}a_{j}^{2}u_{e}/\gamma _{se}^{2}\right] }{\left[
a_{e}^{2}/(\gamma _{e}^{2}\gamma _{sj}^{2})-a_{j}^{2}/(\gamma _{j}^{2}\gamma
_{se}^{2})\right] }\pm \frac{\frac{\gamma _{e}\gamma _{j}}{\gamma
_{se}\gamma _{sj}}a_{j}a_{e}\left[ u_{j}^{2}-2u_{j}u_{e}+u_{e}^{2}\right]
^{1/2}}{\left[ a_{e}^{2}/(\gamma _{e}^{2}\gamma _{sj}^{2})-a_{j}^{2}/(\gamma
_{j}^{2}\gamma _{se}^{2})\right] }  \label{C28}
\end{equation}
with the resonant solution given by
\begin{equation}
v_{w}^{\ast }=\frac{\omega }{k}=\frac{(\gamma _{se}a_{e})\gamma
_{j}u_{j}+(\gamma _{sj}a_{j})\gamma _{e}u_{e}}{\gamma _{j}(\gamma
_{se}a_{e})+\gamma _{e}(\gamma _{sj}a_{j})}~.  \label{C29}
\end{equation}

Inserting the resonant solution (eq.\ C29) into the expression for $\varepsilon $ gives
$$
0.695\leq \varepsilon ^{2}=\frac{\Gamma _{j}^{2}\gamma _{sj}^{4}}{\Gamma
_{e}^{2}\gamma _{se}^{4}}\leq 1.44
$$
where $2.78\leq \Gamma ^{2}\gamma _{s}^{4}\leq 4$. When $a_{j}=a_{e}$ and $
\Gamma _{j}=\Gamma _{e}$, $\varepsilon ^{2}=1$, and the resonant solution is
exact.  The small range on $\varepsilon $ ($0.83\leq \varepsilon \leq 1.2$)
suggests that this solution remains relatively robust for unequal values of
the sound speed and adiabatic index in the jet and external medium. In the
absence of an external flow the resonant solution
$$
v_{w}^{\ast }=\frac{\omega }{k}=\frac{(\gamma _{se}a_{e})\gamma _{j}u_{j}}{
\gamma _{j}(\gamma _{se}a_{e})+(\gamma _{sj}a_{j})}=\frac{\gamma
_{j}\left( M_{j}^{2}-\beta ^{2}\right) ^{1/2}}{\gamma _{j}\left(
M_{j}^{2}-\beta ^{2}\right) ^{1/2}+\left( M_{e}^{2}-\beta ^{2}\right) ^{1/2}}
$$
is equivalent to the form given in Hardee (2000).

The resonant frequencies can be estimated using the large argument forms for
the Bessel and Hankel functions.  In this limit the dispersion relation
becomes
\begin{equation}
\frac{J_{n}^{^{\prime }}(\beta _{j}R)H_{n}^{(1)}(\beta _{e}R)}{J_{n}(\beta
_{j}R)H_{n}^{(1)^{\prime }}(\beta _{e}R)}\approx i\tan (\beta _{j}R-\frac{
2n+1}{4}\pi )=\frac{\chi _{j}}{\chi _{e}}\frac{\beta _{e}}{\beta _{j}}~. \label{C30}
\end{equation}
From the dispersion relation with $Z_{e}+Z_{j}\approx 0$, and $(\chi _{j}/\beta
_{j})(\beta _{e}/\chi _{e})=Z_{j}/Z_{e}\approx \beta _{e}/\beta _{j}\approx
-1$, $\tan [\beta _{j}R-(2n+1)\pi /4]_{\rm Re}\approx 0$ on the real axis.
It follows that
$$
\left| \beta _{j}R\right| \approx \left| \beta _{e}R\right| \approx
(2n+1)\pi /4+m\pi
$$
can be used to obtain an estimate for the resonant frequencies from $\left|
\beta _{e}R\right| \approx (2n+1)\pi /4+m\pi$, with result that the
resonant frequencies are given by
\begin{equation}
\frac{\omega_{nm}R}{a_{e}}  \approx \frac{\omega _{nm}^{\ast
  }R}{a_{e}} \equiv \frac{(2n+1)\pi /4+m\pi }{\gamma
_{e}\left[ \left( 1- u_{e}/v_w^{\ast} \right) ^{2}-\left(a_{e}/v_w^{\ast}
-u_{e}a_{e}/c^{2}\right) ^{2}\right] ^{1/2}}~.  \label{C31}
\end{equation}
In the absence of external
flow, $u_{e}=0$, and for $u_{j}>>a_{e}$ and  $1>>(ka_{e}/\omega )^{2}$ this
expression reduces to the form given in Hardee (2000). 

When $\gamma
_{j}(\gamma _{se}a_{e})>>\gamma _{e}(\gamma _{sj}a_{j})$, the resonant wave
speed becomes $v_w^{\ast}\approx u_{j}$, $u_{e}/v_w^{\ast} \approx
u_{e}/u_{j}$ and provided $u_{e}<<u_{j}$ and $a_{e}<<u_{j}$, the resonant
frequency increases with increasing $u_{e}/u_{j}$ and $a_{e}/u_{j}$
as
\begin{equation}
\frac{\omega _{nm}^{\ast }R}{a_{e}}\approx \frac{(2n+1)\pi /4+m\pi }{\left[
1-2u_{e}/u_{j}(1-a_{e}^{2}/c^{2})-(a_{e}^{2}-u_e^2)/u_{j}^{2}\right] ^{1/2}}~.
\label{C32}
\end{equation}
In general, the resonant frequency $\omega _{nm}^{\ast }\longrightarrow
\infty $ as $\left( 1-u_{e}/v_w^{\ast} \right) ^{2}-\left(
a_{e}/v_w^{\ast} -u_{e}a_{e}/c^{2}\right) ^{2}\longrightarrow 0$. An
equivalent condition for $\left( 1-u_{e}/v_w^{\ast} \right) ^{2}-\left(
a_{e}/v_w^{\ast} -u_{e}a_{e}/c^{2}\right) ^{2}=0$ is
\begin{equation}
\frac{u_{j}-u_{e}}{1-u_{j}u_{e}/c^{2}}=\frac{a_{j}+a_{e}}{
1+a_{j}a_{e}/c^{2}}~,  \label{C33}
\end{equation}
and the resonance moves to higher frequencies with $\omega
_{nm}^{\ast }\longrightarrow \infty $ when the``shear'' speed drops below a
``surface'' sound speed.

The behavior of the growth rate at resonance also can be found using the
large argument forms for the Bessel and Hankel functions.  In this limit
the reflectance can be written as
\begin{equation}
\mathcal{R}=\frac{(Z_{e}-Z_{j})}{(Z_{e}+Z_{j})}=\frac{J_{n}(\beta
_{j}R)H_{n}^{(1)^{\prime }}(\beta _{e}R)-J_{n}^{^{\prime }}(\beta
_{j}R)H_{n}^{(1)}(\beta _{e}R)}{J_{n}(\beta _{j}R)H_{n}^{(1)^{\prime
}}(\beta _{e}R)+J_{n}^{^{\prime }}(\beta _{j}R)H_{n}^{(1)}(\beta _{e}R)}
\approx \exp [-2i(\beta _{j}R-\frac{2n+1}{4}\pi )]~,  \label{C34}
\end{equation}
and
\begin{equation}
\beta _{j}R-\frac{2n+1}{4}\pi \approx \frac{i}{2}\ln \left| \mathcal{R}%
\right| -\frac{\phi }{2} \label{C35}
\end{equation}
where $\mathcal{R\equiv }\left| \mathcal{R}\right| e^{i\phi }$. It follows
that
\begin{equation}
(\beta _{j}R)_{I}\approx \frac{1}{2}\ln \left| \mathcal{R}\right| \label{C36}
\end{equation}
and since typically at resonance, $\left| \omega -k_{R}u_{j}\right|
/a_{j}>\left| k_{R}-\omega u_{j}/c^{2}\right|$ I can approximate $\beta
_{j}$ by
\begin{equation}
\beta _{j}\equiv \beta _{j}^{R}+i\beta _{j}^{I}\approx \gamma _{j}\left[ 
\frac{\left( \omega -k_{R}u_{j}\right) }{a_{j}}-ik_{I}\left( \frac{u_{j}}{
a_{j}}\right) \right]~.  \label{C37}
\end{equation}
It follows that
\begin{equation}
(\beta _{j}R)_{I}\approx -\gamma _{j}\left( \frac{u_{j}}{a_{j}}\right)
k_{I}R~, \label{C38}
\end{equation}
and
\begin{equation}
k_{I}R\approx -\frac{1}{2}\frac{a_{j}}{\gamma _{j}u_{j}}\ln \left| \mathcal{R
}\right|~. \label{C39}
\end{equation}

At resonance
\begin{equation}
\mathcal{R}=\frac{(Z_{e}-Z_{j})}{(Z_{e}+Z_{j})}=\frac{\beta _{j}-\beta
_{e}(\chi _{j}/\chi _{e})}{\beta _{j}+\beta _{e}(\chi _{j}/\chi _{e})}
\approx \frac{\beta _{j}-\beta _{e}}{\beta _{j}+\beta _{e}}\approx \frac{
-2\beta _{e}}{\beta _{j}+\beta _{e}}  \label{C40}
\end{equation}
and
\begin{equation}
\left| \mathcal{R}\right| \approx \left| \frac{\beta _{j}-\beta _{e}}{\beta
_{j}+\beta _{e}}\right| \approx \left[ \frac{\left( -2\beta _{e}^{R}\right)
^{2}+\left( \beta _{j}^{I}-\beta _{e}^{I}\right) ^{2}}{\left( \beta
_{j}^{I}+\beta _{e}^{I}\right) ^{2}}\right] ^{1/2}  \label{C41}
\end{equation}
where I have used $\left( \beta _{j}^{R}-\beta _{e}^{R}\right) \approx
-2\beta _{e}^{R}$ from the resonance condition on the real axis. It follows
that
\begin{equation}
\left| \mathcal{R}\right| \approx \left[ \frac{4\gamma _{e}^{2}\frac{\left(
\omega -k_{R}u_{e}\right) ^{2}}{a_{e}^{2}}+k_{I}^{2}\left[ \left( \gamma _{j}
\frac{u_{j}}{a_{j}}-\gamma _{e}\frac{u_{e}}{a_{e}}\right) -\frac{\gamma
_{e}a_{e}}{\omega -k_{R}u_{e}}k_{R}\right] ^{2}}{k_{I}^{2}\left[ \left(
\gamma _{j}\frac{u_{j}}{a_{j}}+\gamma _{e}\frac{u_{e}}{a_{e}}\right) +\frac{
\gamma _{e}a_{e}}{\omega -k_{R}u_{e}}k_{R}\right] ^{2}}\right] ^{1/2}
\label{C42}
\end{equation}
where I have used 
$$
\beta _{e}\equiv \beta _{e}^{R}+i\beta _{e}^{I}\approx \gamma _{e}\left[ 
\frac{\left( \omega -k_{R}u_{e}\right) }{a_{e}}-ik_{I}\left( \frac{u_{e}}{
a_{e}}+\frac{a_{e}}{\omega -k_{R}u_{e}}k_{R}\right) \right]~.
$$

If I assume that $\gamma _{j}(\gamma _{se}a_{e})>>\gamma _{e}(\gamma
_{sj}a_{j})$, with resonant wave speed $\omega /k_{R}\approx u_{j}$ and $
u_{e}/u_{j}<<1$ then
\begin{equation}
\left| \mathcal{R}\right| \approx \left[ \frac{4\left( \frac{\omega
_{nm}^{\ast }R}{a_{e}}\right) ^{2}\left( 1-2u_{e}/u_{j}\right)
+k_{I}^{2}R^{2}\left[ \gamma _{j}\frac{u_{j}}{a_{j}}-\frac{a_{e}}{u_{j}}
\left( 1+u_{e}/u_{j}\right) \right] ^{2}}{k_{I}^{2}R^{2}\left[ \gamma _{j}
\frac{u_{j}}{a_{j}}+\frac{a_{e}}{u_{j}}\left( 1+u_{e}/u_{j}\right) \right]
^{2}}\right] ^{1/2}~,  \label{C43}
\end{equation}
and since $k_{I}R\approx -\left( a_{j}/2\gamma _{j}u_{j}\right) \ln \left| 
\mathcal{R}\right|$
\begin{equation}
\left| \mathcal{R}\right| \approx \left[ \frac{4\left( \frac{\omega
_{nm}^{\ast }R}{a_{e}}\right) ^{2}\left( 1-2u_{e}/u_{j}\right) +\left[ \ln
\left| \mathcal{R}\right| /2\right] ^{2}}{\left[ \ln \left| \mathcal{R}
\right| /2\right] ^{2}}\right] ^{1/2}~.  \label{C44}
\end{equation}
From equation (C32)
$$
\left( \frac{\omega _{nm}^{\ast }R}{a_{e}}\right) ^{2}\left(
1-2u_{e}/u_{j}\right) \approx \frac{\left( 1-2u_{e}/u_{j}\right) }{\left[
1-2\left( u_{e}/u_{j}\right) (1-a_{e}^{2}/c^{2})-a_{e}^{2}/u_{j}^{2}\right] }
\left[ (2n+1)\pi /4+m\pi \right] ^{2}
$$
and if say $u_{e}=0$, then
\begin{equation}
\left( \left| \mathcal{R}\right| ^{2}-1\right) ^{1/2}\ln \left| \mathcal{R}
\right| \approx 4\left[ \frac{1}{1-a_{e}^{2}/u_{j}^{2}}\right] ^{1/2}\left[
(2n+1)\pi /4+m\pi \right]~.  
\label{C45}
\end{equation}
Formally $\left| \mathcal{R}
\right| \longrightarrow \infty $ as $\omega _{nm}^{\ast }\longrightarrow
\infty $ when the jet speed drops below the ``surface'' sound speed given by
equation (C33). This result applies only to the surface modes and not to the
body modes as, in the fluid limit, the body modes do not exist when the jet
speed drops below the jet sound speed, see equation (C17). On the other hand, if
say, $a_{e}^{2}/u_{j}^{2}<<1$, then
\begin{equation}
\left( \left| \mathcal{R}\right| ^{2}-1\right) ^{1/2}\ln \left| \mathcal{R}
\right| \approx 4\left[ (2n+1)\pi /4+m\pi \right]~.  
\label{C46}
\end{equation}
Formally $
\left| \mathcal{R}\right| \approx $ constant as $\omega _{nm}^{\ast
}\longrightarrow \infty $ when the wind speed becomes comparable to the jet
speed, $u_{e}\lesssim u_{j}$, as must be the case for the velocity shear
driven Kelvin-Helmholtz instability.

\vspace{-0.5cm}
\subsubsection{The Magnetic Limit (Alfv\'{e}n speed $\gg$ sound speed)}

For the magnetic limit in which magnetic pressure dominates gas pressure
\begin{equation} \eqnum{C47a}
Z_{e}=\gamma _{e}\gamma _{Ae}^{2}W_{e}v_{Ae}\left( \varpi
_{e}^{2}-\kappa _{e}^{2}v_{Ae}^{2}\right) ^{1/2}~,  \label{C47a}
\end{equation}
and
\begin{equation} \eqnum{C47b}
Z_{j}=\gamma _{j}\gamma _{Aj}^{2}W_{j}v_{Aj}\left( \varpi _{j}^{2}-\kappa
_{j}^{2}v_{Aj}^{2}\right) ^{1/2}~,  \label{C47b}
\end{equation}
\setcounter{equation}{47}
A necessary condition for resonance is $\left( \varpi
_{e}^{2}-\kappa _{e}^{2}v_{Ae}^{2}\right) >0$ and $\left( \varpi
_{j}^{2}-\kappa _{j}^{2}v_{Aj}^{2}\right) <0$ on the real axis with result that $
Z_{e}+Z_{j}=0 $ when
\begin{equation}
\frac{u_{j}-v_{Aj}}{1-u_{j}v_{Aj}/c^{2}}>\frac{\omega }{k}>\frac{u_{e}+v_{Ae}
}{1+u_{e}v_{Ae}/c^{2}}~.  \label{C48}
\end{equation}
It follows that the resonance only exists when
\begin{equation}
\frac{u_{j}-u_{e}}{1-u_{j}u_{e}/c^{2}}>\frac{v_{Aj}+v_{Ae}}{
1+v_{Aj}v_{Ae}/c^{2}}~.  \label{C49}
\end{equation}
This result is identical in form to the sonic case with sound speeds
replaced by Alfv\'{e}n wave speeds. 

The resonant solution for the
real part of the phase velocity is obtained from
\begin{equation}
Z_{j}^{2}=\gamma _{j}^{2}W_{j}^{2}V_{Aj}^{2}\left[ \varpi
_{j}^{2}-\left( k^{2}-\omega ^{2}/c^{2}\right) V_{Aj}^{2}/\gamma _{j}^{2}
\right] =Z_{e}^{2}=\gamma _{e}^{2}W_{e}^{2}V_{Ae}^{2}\left[ \varpi
_{e}^{2}-\left( k^{2}-\omega ^{2}/c^{2}\right) V_{Ae}^{2}/\gamma _{e}^{2}
\right]~,  \label{C50}
\end{equation}
where I have used $\gamma ^{2}\gamma _{A}^{2}\left( \varpi ^{2}-\kappa
^{2}v_{A}^{2}\right) =\gamma ^{2}\left[ \varpi ^{2}-\left( k^{2}-\omega
^{2}/c^{2}\right) V_{A}^{2}/\gamma ^{2}\right]$, and recall that $
v_{A}^{2}=V_{A}^{2}/\gamma _{A}^{2}$. The resulting quadratic equation can
be written in the form
\doublespace
\begin{equation}
\begin{array}{c}
\left[ \gamma _{j}^{2}W_{j}^{2}V_{Aj}^{2}-\gamma _{e}^{2}W_{e}^{2}V_{Ae}^{2}
\right] \left( \frac{\omega }{k}\right) ^{2}-2\left[ \gamma
_{j}^{2}W_{j}^{2}V_{Aj}^{2}u_{j}-\gamma _{e}^{2}W_{e}^{2}V_{Ae}^{2}u_{e}
\right] \left( \frac{\omega }{k}\right) \\ +\left[ \gamma
_{j}^{2}W_{j}^{2}V_{Aj}^{2}u_{j}^{2}-\gamma
_{e}^{2}W_{e}^{2}V_{Ae}^{2}u_{e}^{2}\right] =0
\end{array} \label{C51}
\end{equation}
\baselineskip 12pt
where I have used
$$
\gamma _{j}^{2}W_{j}^{2}V_{Aj}^{2}\left( k^{2}-\omega ^{2}/c^{2}\right)
V_{Aj}^{2}/\gamma _{j}^{2}=\gamma _{e}^{2}W_{e}^{2}V_{Ae}^{2}\left(
k^{2}-\omega ^{2}/c^{2}\right) V_{Ae}^{2}/\gamma _{e}^{2}
$$
because pressure balance in the
magnetically dominated case requires $W_{j}V_{Aj}^{2}=W_{e}V_{Ae}^{2}$. The
solutions are given by
\begin{equation}
\frac{\omega }{k}=\frac{\gamma _{j}^{2}W_{j}^{2}V_{Aj}^{2}u_{j}-\gamma
_{e}^{2}W_{e}^{2}V_{Ae}^{2}u_{e}\pm \gamma _{j}\gamma
_{e}W_{j}W_{e}V_{Aj}V_{Ae}\left( u_{j}-u_{e}\right) }{\gamma
_{j}^{2}W_{j}^{2}V_{Aj}^{2}-\gamma _{e}^{2}W_{e}^{2}V_{Ae}^{2}}~,  \label{C52}
\end{equation}
and the resonant solution becomes
\begin{equation}
v_{w}^{\ast }=\frac{\omega }{k}=\frac{\gamma _{j}W_{j}V_{Aj}u_{j}+\gamma
_{e}W_{e}V_{Ae}u_{e}}{\gamma _{j}W_{j}V_{Aj}+\gamma _{e}W_{e}V_{Ae}}=\frac{
\left( \gamma _{Ae}v_{Ae}\right) \gamma _{j}u_{j}+\left( \gamma
_{Aj}v_{Aj}\right) \gamma _{e}u_{e}}{\gamma _{j}\left( \gamma
_{Ae}v_{Ae}\right) +\gamma _{e}\left( \gamma _{Aj}v_{Aj}\right) }
\label{C53}
\end{equation}
where I have used $WV_{A}=WV_{A}^{2}/\left( \gamma _{A}v_{A}\right)$, and $
W_{j}V_{Aj}^{2}=W_{e}V_{Ae}^{2}$. This resonant solution has the same form
as the sonic case with sound speeds and sonic Lorentz factors replaced by
Alfv\'{e}n wave speeds and Alfv\'{e}nic Lorentz factors.

As in the sonic case the resonant frequencies are found from
$$
\left| \beta _{e}R\right| \approx (2n+1)\pi /4+m\pi
$$
with result that the resonant frequencies are given by
\begin{equation}
\frac{\omega_{nm}R}{v_{Ae}}\approx\frac{\omega _{nm}^{\ast
  }R}{v_{Ae}} \equiv \frac{(2n+1)\pi /4+m\pi }{\gamma
_{e}\left[ \left( 1-u_{e}/v_w^{\ast} \right) ^{2}-\left( v_{Ae}/v_w^{\ast}
-u_{e}v_{Ae}/c^{2}\right) ^{2}\right] ^{1/2}}~.  \label{C54}
\end{equation}
When $\gamma _{j}\left(
\gamma _{Ae}v_{Ae}\right) >>\gamma _{e}\left( \gamma _{Aj}v_{Aj}\right) $
the resonant wave speed becomes $v_w^{\ast}\approx u_{j}$ and $
u_{e}/v_w^{\ast} \approx u_{e}/u_{j}$, and provided $u_{e}<<u_{j}$ and $
v_{Ae}<<u_{j}$ the resonant frequency increases with increasing $u_{e}/u_{j}$
and $v_{Ae}/u_{j}$ as
\begin{equation}
\frac{\omega _{nm}^{\ast }R}{v_{Ae}}\approx \frac{(2n+1)\pi /4+m\pi }{\left[
1-2\left( u_{e}/u_{j}\right) (1-v_{Ae}^{2}/c^{2})-(v_{Ae}^{2}-u_e^2)/u_{j}^{2}\right]
^{1/2}}~.  \label{C55}
\end{equation}
Here the resonant frequency $\omega _{nm}^{\ast }\longrightarrow \infty $ as 
$\left( 1-u_{e}/v_w^{\ast} \right) ^{2}-\left( v_{Ae}/v_w^{\ast}
-u_{e}v_{Ae}/c^{2}\right) ^{2}\longrightarrow 0$. An equivalent condition
for $\left( 1- u_{e}/v_w^{\ast} \right) ^{2}-\left( v_{Ae}/v_w^{\ast}
-u_{e}v_{Ae}/c^{2}\right) ^{2}=0$ is
\begin{equation}
\frac{u_{j}-u_{e}}{1-u_{j}u_{e}/c^{2}}=\frac{v_{Aj}+v_{Ae}}{
1+v_{Aj}v_{Ae}/c^{2}}~,  \label{C56}
\end{equation}
and the resonance moves to higher frequencies as the ``shear'' speed becomes
trans-Alfv\'{e}nic.

The behavior of the growth rate at resonance proceeds in the same manner as
for the fluid limit but with sound speeds replaced by Alfv\'{e}n wave
speeds. The resonant growth rate is now given by
\begin{equation}
k_{I}R\approx -\frac{1}{2}\frac{v_{Aj}}{\gamma _{j}u_{j}}\ln \left| \mathcal{
R}\right|~.  \label{C57}
\end{equation}
If I assume that $\gamma _{j}(\gamma _{Ae}v_{Ae})>>\gamma _{e}(\gamma
_{Aj}v_{Aj})$, with resonant wave speed $\omega /k_{R}\approx u_{j}$ and $
u_{e}/u_{j}<<1$ then
\begin{equation}
\left| \mathcal{R}\right| \approx \left[ \frac{4\left( \frac{\omega
_{nm}^{\ast }R}{v_{Ae}}\right) ^{2}\left( 1-2u_{e}/u_{j}\right)
+k_{I}^{2}R^{2}\left[ \gamma _{j}\frac{u_{j}}{v_{Aj}}-\frac{v_{Ae}}{u_{j}}
\left( 1+u_{e}/u_{j}\right) \right] ^{2}}{k_{I}^{2}R^{2}\left[ \gamma _{j}
\frac{u_{j}}{v_{Aj}}+\frac{v_{Ae}}{u_{j}}\left( 1+u_{e}/u_{j}\right) \right]
^{2}}\right] ^{1/2}~,  \label{C58}
\end{equation}
From equation (C54)
$$
\left( \frac{\omega _{nm}^{\ast }R}{v_{Ae}}\right) ^{2}\left(
1-2u_{e}/u_{j}\right) \approx \frac{\left( 1-2u_{e}/u_{j}\right) }{\left[
1-2\left( u_{e}/u_{j}\right) (1-v_{Ae}^{2}/c^{2})-(v_{Ae}^{2}-u_e^2)/u_{j}^{2}\right]
}\left[ (2n+1)\pi /4+m\pi \right] ^{2}
$$
and if say $u_{e}=0$, then
\begin{equation}
\left( \left| \mathcal{R}\right| ^{2}-1\right) ^{1/2}\ln \left| \mathcal{R}
\right| \approx 4\left[ \frac{1}{1-v_{Ae}^{2}/u_{j}^{2}}\right] ^{1/2}\left[
(2n+1)\pi /4+m\pi \right]  \label{C59}
\end{equation}
and $\left| \mathcal{R}\right| $ increases as $\omega _{nm}^{\ast }$
increases when the jet speed decreases.  However, when the shear speed
drops below the ``surface'' Alfv\'{e}n speed, see equations (C11 \& C12) the jet is
stable.  This result is quite different from the fluid limit where the jet
remains unstable when the shear speed drops below the ``surface'' sound
speed. If I insert
$$
u_{j}-u_{e}=\frac{1-u_{j}u_{e}/c^{2}}{1+v_{Aj}v_{Ae}/c^{2}}\left(
v_{Aj}+v_{Ae}\right)~.
$$
from equation (C56) into equation (C12), it follows that the jet will be unstable
when resonance disappears only when
\begin{equation}
\gamma _{j}^{2}\gamma _{e}^{2}\left( 1-u_{j}u_{e}/c^{2}\right) ^{2}>2\gamma
_{Aj}^{2}\gamma _{Ae}^{2}\frac{v_{Ae}^{2}+v_{Aj}^{2}}{\left(
v_{Aj}+v_{Ae}\right) ^{2}}\left( 1+v_{Aj}v_{Ae}/c^{2}\right) ^{2}~.
\label{C60}
\end{equation}
where I have used $\left( v_{Ae}^{2}+v_{Aj}^{2}\right) =\left( W_{j}/\gamma
_{Ae}^{2}+W_{e}/\gamma _{Aj}^{2}\right) \left[ \left(
B_{j}^{2}+B_{e}^{2}\right) /\left( 4\pi W_{j}W_{e}\right) \right] $ as $
B_{e}=B_{j}$ from magnetic pressure balance. Formally $\left| \mathcal{R}
\right| \longrightarrow \infty $ as $\omega _{nm}^{\ast }\longrightarrow
\infty $ only for jet Lorentz factors greatly in excess of the Alfv\'{e}nic Lorentz factor.

\vspace{-0.5cm}
\subsection{Wave modes at high frequency}

To obtain the behavior of wave modes at high frequency I begin with the
dispersion relation written in the form
\begin{equation}
\frac{\beta _{j}}{\beta _{e}}\frac{\chi _{e}}{\chi _{j}}=\frac{J_{n}(\beta
_{j}R)}{J_{n}^{^{\prime }}(\beta _{j}R)}\frac{H_{n}^{(1)^{\prime }}(\beta
_{e}R)}{H_{n}^{(1)}(\beta _{e}R)}=\frac{J_{n}(\beta _{j}R)}{\mp J_{n\pm
1}(\beta _{j}R)\pm \frac{n}{\beta _{j}R}J_{n}(\beta _{j}R)}\frac{
H_{n-1}^{(1)}(\beta _{e}R)-\frac{n}{\beta _{e}R}H_{n}^{(1)}(\beta _{e}R)}{
H_{n}^{(1)}(\beta _{e}R)}  \label{C61}
\end{equation}
and assume a large argument in the Hankel function with 
$H_{n}^{(1)}(\beta _{e}R)\approx \exp i\left[ \beta _{e}R-\left( 2n+1\right)
\pi /4\right] $ and a small argument $\beta _{j}R<<1$ in the Bessel
function to write
$$
\frac{\beta _{j}R}{\beta _{e}R}\frac{\chi _{e}}{\chi _{j}} \approx \left\{
\begin{array}{cc}
\frac{J_{0}(\beta _{j}R)}{-J_{1}(\beta _{j}R)}e^{-i\pi /2} & n=0 \\ 
\frac{\beta _{j}R}{n}e^{-i\pi /2} & n>0
\end{array} \right. ~.
$$
The small arguement form for the Bessel function gives $J_{0}(\beta
_{j}R)/J_{1}(\beta _{j}R)\approx 2/\beta _{j}R$ with result that the
dispersion relation becomes
\begin{equation}
\beta _{j}R\approx \left\{
\begin{array}{cc}
\left[ -\left( \frac{\chi _{j}}{\chi _{e}}\beta _{e}R\right) e^{-i\pi /2}
\right] ^{1/2} & n=0 \\ 
\frac{\beta _{j}R}{n}\left( \frac{\chi _{j}}{\chi _{e}}\beta _{e}R\right)
e^{-i\pi /2} & n>0
\end{array} \right. ~.
\label{C62}
\end{equation}
At high frequency and large wavenumber $\chi _{j}$ and $\chi _{e}$, are
proportional to $k^{2}$, $\beta _{j}$ and $\beta _{e}$ are proportional
to $k$, and $\beta _{j}R=\zeta _{j}kR\propto \left( kR\right) ^{1/2}$ for $
n=0$.  Thus, the internal solutions in the high frequency and large
wavenumber limit are given by $\beta _{j}R\simeq 0$ and are found from
\begin{equation}
\left[ \left( ku_{j}-\omega \right) ^{2}-\left( k-\omega u_{j}/c^{2}\right)
^{2}a_{j}^{2}\right] \left[ \left( ku_{j}-\omega \right) ^{2}-\left(
k-\omega u_{j}/c^{2}\right) ^{2}v_{Aj}^{2}\right] \approx 0  \label{C63}
\end{equation}
and it follows that
\begin{equation} \eqnum{C64a}
\frac{\omega }{k}\approx \frac{u_{j}\pm a_{j}}{1\pm u_{j}a_{j}/c^{2}}~,  \label{C64a}
\end{equation}
or
\begin{equation} \eqnum{C64b}
\frac{\omega }{k}\approx \frac{u_{j}\pm v_{Aj}}{1\pm u_{j}v_{Aj}/c^{2}}~. \label{C64b}
\end{equation}
\setcounter{equation}{64}

\end{document}